\pdfoutput=1

\documentclass[11pt,twoside,a4paper,cmspaper,final,collab]{cms-tdr}
\def\svnVersion{168100}\def\cmsCernNo{2012-244}\def\cmsCernDate{\today}\def\cmsMessage{Submitted to Physics Letters B}

\begin{document}\cmsNoteHeader{EXO-11-099}

\hyphenation{had-ron-i-za-tion}
\hyphenation{cal-or-i-me-ter}
\hyphenation{de-vices}

\RCS$Revision: 154532 $
\RCS$HeadURL: svn+ssh://svn.cern.ch/reps/tdr2/papers/EXO-11-099/trunk/EXO-11-099.tex $
\RCS$Id: EXO-11-099.tex 154532 2012-10-24 14:56:31Z vorobiev $
\providecommand{\Mfit}{\ensuremath{M_\text{fit}\xspace}}
\newlength\cmsFigWidth
\ifthenelse{\boolean{cms@external}}{\setlength\cmsFigWidth{0.95\columnwidth}}{\setlength\cmsFigWidth{0.6\textwidth}}
\ifthenelse{\boolean{cms@external}}{\providecommand{\cmsLeft}{top}}{\providecommand{\cmsLeft}{left}}
\ifthenelse{\boolean{cms@external}}{\providecommand{\cmsRight}{bottom}}{\providecommand{\cmsRight}{right}}
\providecommand{\CL[1]}{\ensuremath{\text{CL}_\text{#1}}\xspace}
\cmsNoteHeader{EXO-11-099} 
\title{Search for pair produced fourth-generation up-type quarks in pp collisions at $\sqrt{s}= $ 7\TeV with a lepton in the final state}

\date{\today}

\abstract{
The results of a search for the pair-production of a fourth-generation
up-type quark ($\PQtpr$) in proton-proton collisions at $\sqrt{s} = 7$\TeV
are presented, using data corresponding to an integrated luminosity of about
5.0\fbinv collected by the Compact Muon Solenoid experiment at the LHC.
The $\PQtpr$ quark is assumed
to decay exclusively to a $\PW$ boson and a $\cPqb$ quark. Events with
a single isolated electron or muon, missing transverse momentum, and at least four
hadronic jets, of which at least one must be identified as a $\cPqb$ jet, are
selected. No significant excess of events over standard model expectations is observed.
Upper limits for the $\PQtpr\PAQtpr$ production cross section at 95\% confidence
level are set as a function of $\PQtpr$ mass, and $\PQtpr$-quark production
for masses below 570\GeV is excluded. The search is equally sensitive to nonchiral
heavy quarks decaying to $\PW\cPqb$. In this case, the results can be interpreted
as upper limits on the production cross section times the branching fraction to
$\PW\cPqb$.
}

\hypersetup{%
pdfauthor={CMS Collaboration},%
pdftitle={Search for pair produced fourth-generation up-type quarks in pp collisions at sqrt(s) = 7 TeV with a lepton in the final state},%
pdfsubject={CMS},%
pdfkeywords={CMS, physics}}

\maketitle 

\section{Introduction}

The standard model (SM) of particle physics includes three generations of
fermions~\cite{Glashow:1961tr,PhysRevLett.19.1264,salam}. But the
possibility of a sequential fourth generation of fermions is not ruled out
by precision electroweak measurements (see, e.g.~\cite{novikov}). If the
recently observed 125\GeV boson~\cite{Atlas_Higgs,CMS_Higgs}
turns out to be the SM Higgs boson, the existence of a sequential fourth
generation would be disfavoured (see, e.g.~\cite{Eberhardt_2012}).

However, many models of physics beyond the SM, such as theories with an
extended Higgs sector (see, e.g.~\cite{Chen_He}), can still incorporate a
sequential fourth generation of fermions if the 125\GeV boson turns out to
be one of the predicted extended Higgs bosons. Furthermore, other models
predict (see, e.g.~\cite{beautiful,Schmaltz200340}) the existence of a heavy,
nonchiral, vector-like quark whose left- and right-handed components
transform in the same way under the symmetry group of the theory, as
a partner to the top quark. This particle would cancel the divergent
corrections of \cPqt-quark loops to the Higgs boson mass.
The production of such a nonsequential quark would give rise to the same
final-state signature described below in the search for a sequential up-type
quark (denoted as $\PQtpr$ in this paper), and thus the results of
this search are also relevant for it.

For a fourth-generation up-type quark, the mass splitting between the
$\PQtpr$ quark and the corresponding down-type $\PQbpr$ quark
is favoured to be smaller than the mass of the $\PW$
boson~\cite{Kribs, Soni, Eberhardt}. In this case, the $\PQtpr$ quark
cannot decay to $\PW\PQbpr$. Assuming that the pattern of quark mixing
observed in the CKM matrix extends to the fourth generation, the dominant
$\PQtpr$-quark decay mode would then be $\PQtpr\to\PW\cPqb$,
and the lifetime of the $\PQtpr$ would be short, similar to that of
the top quark.

It is interesting to note that the coupling to the Higgs field of a
fourth-generation quark with a mass above about 550\GeV becomes large, its
weak interactions start to become comparable to its strong interactions, and
perturbative calculations begin to fail~\cite{Chanowitz}.
However, such effects are still small at the beginning of this mass
range, as has been shown for the case of CKM mixing~\cite{Chanowitz_CKM}.

We present the results of a search for the strong production of $\PQtpr\PAQtpr$
quark pairs, with $\PQtpr$ decaying into $\PWp\cPqb$ and
$\PAQtpr$ to $\PWm\cPaqb$, in proton-proton collisions at
$\sqrt{s} = 7$\TeV using the Compact Muon Solenoid (CMS) detector. The search
strategy requires that one of the $\PW$ bosons decays to leptons
($\Pe\nu$ or $\mu\nu$) and the other to a quark-antiquark pair.
The branching fraction into these final states
is about 15\% for each lepton flavour. We select events with a single charged lepton,
missing transverse momentum, and at least four jets with high transverse momenta ($\pt$).

Previous searches for $\PQtpr$ quarks in this final state give lower limits for
the mass of the $\PQtpr$ quark of 358\GeV~\cite{D0,CDF} at the Tevatron and
404\GeV~\cite{ATLAS_1lepton} at the Large Hadron Collider (LHC).

There are SM processes that give rise to the same signature, most notably
$\ttbar$ and $\PW$+jets production. The present search considers
a $\PQtpr$ quark with a mass larger than the SM $\mathrm{t}$ quark.
We utilize two variables to distinguish between signal and background.
The first is $H_\mathrm{T}$, defined as the scalar sum of the transverse momenta of
the lepton, the missing transverse momentum, and the four jets from the decay
of the $\PQtpr$ and $\PAQtpr$ quarks.
The second variable is the $\PQtpr$-quark mass $M_{\mathrm{fit}}$, obtained from
a kinematic fit of each event to the process
$\PQtpr\PAQtpr\to \PW\cPqb\PW\cPaqb\to \ell\nu\cPqb\cPq\cPaq'\cPaqb$ .
We use the two-dimensional distribution of $H_\mathrm{T}$ versus
$M_{\text{fit}}$
to test for the presence of a signal for
$\mathrm{t'\bar{t'}}$ production in the data.

We categorize events according to the flavour of the lepton.
Events with an identified electron (muon) are classified as $\Pe$+jets
($\mu$+jets) events. The analysis procedures for the two channels are
kept as similar as possible, with small differences mainly driven by the
different trigger conditions. Finally, a combined statistical analysis of
both channels is performed and upper limits for the $\PQtpr\PAQtpr$
pair production cross section and a lower limit on the $\PQtpr$-quark mass
are derived.

\section{CMS detector and data samples}
\label{sec:data}

The CMS experiment uses the following coordinate system. The $z$ axis coincides
with the axis of symmetry of the detector, and is oriented in
the anticlockwise proton beam direction. The $x$ axis points towards the centre
of the LHC ring and the $y$ axis points up.
The polar angle $\theta$ is defined with respect to the positive $z$
axis, and $\phi$ is the azimuthal angle. The transverse momentum
of a particle or jet is defined as its momentum times $\sin{\theta}$, and pseudorapidity
is ${\eta = -\ln[\tan(\theta / 2)]}$.

The characteristic feature of the CMS detector is the superconducting
solenoid, 6\unit{m} in diameter and 13\unit{m} in length, which provides an axial
magnetic field of 3.8\unit{T}. Inside the solenoid are a multi-layered silicon pixel
and strip tracker covering $|\eta| < 2.5$ to measure the trajectories of charged
particles, an electromagnetic calorimeter (ECAL) made of lead tungstate crystals
and covering $|\eta|<3.0$, a preshower detector covering $1.65 < |\eta| < 2.6$
to measure electrons and photons, and a hadronic calorimeter (HCAL) made of
brass and scintillators covering $|\eta| < 3.0$ to measure jets. Muons are
identified using gas-ionization detectors embedded in the steel return yoke of
the solenoid and covering $|\eta| < 2.4$. Extensive forward calorimetry
complements the coverage provided by the barrel and endcap detectors.
The CMS detector is nearly hermetic, allowing the measurement of the transverse
momentum carried by undetected particles. A detailed description of the CMS detector
can be found elsewhere~\cite{CMS}.

We use data collected in 2011, corresponding to an integrated luminosity
of 5.0\fbinv for the \Pe+jets channel and 4.9\fbinv
for the $\mu$+jets channel. The triggers for the $\Pe$+jets data
required at least one electron candidate with a $\pt$ threshold that
varied between 25 and 32\GeV according to the average instantaneous luminosity.
When the LHC instantaneous luminosity increased, three central jets with
$\pt>30$\GeV and $|\eta|<2.6$ were also required. The triggers for the $\Pgm$+jets
channel required at least one muon candidate with a $\pt$ threshold that varied
between 30 and 40\GeV. No requirements were made on jets in the triggers
for the $\Pgm$+jets events.

We model the $\PQtpr\PAQtpr$ signal and SM background processes using Monte
Carlo (MC) simulations.  The $\PQtpr\PAQtpr$ signal events are generated for
$\PQtpr$ masses from 400 to 625\GeV in 25\GeV steps. The following SM background
processes are simulated: $\ttbar$ production; single-$\cPqt$-quark production
via the $\cPqt\PW$, $s$-channel, and $t$-channel processes; single- and
double-boson production ($\PW$+jets, $\cPZ$+jets, $\PW\PW$,
$\PW\cPZ$, and $\cPZ\cPZ$).
All of these processes, except the dominant $\ttbar$ production, are collectively
referred to as electroweak (EW) background.

The single-\cPqt-quark production is simulated with the \POWHEG event
generator~\cite{powheg1,powheg2,powheg3}.
All other processes are simulated with the
\textsc{MadEvent}/\MADGRAPH\cite{madgraph} programs.
The \PYTHIA program~\cite{pythia} is then used to simulate additional
radiation and the fragmentation and hadronization of the quarks and gluons
into jets.
The generated events are processed through the CMS detector simulation
based on \GEANTfour~\cite{geant4}.
Up to 20 minimum-bias events, generated with
\PYTHIA, are superimposed on the hard-scattering events to simulate
multiple $\mathrm{pp}$ interactions within the same beam crossing. The
MC-simulated events are weighted to reproduce the distribution of the number
of vertices per event in the data (the average number of vertices per event
is 8).

The simulated samples for the $\PQtpr\PAQtpr$ signal correspond to an
integrated luminosity of between 100 and 2500\fbinv for each value of
$\PQtpr$-quark mass. Samples for the background processes giving
the largest contributions correspond to 22\fbinv for the $\ttbar$ sample
and 2.5\fbinv for \PW+jets.

\section{Event reconstruction}

Events are reconstructed using a particle-flow
algorithm~\cite{particleflow,particleflow1,particleflow2}.
The particle-flow event reconstruction consists in reconstructing and
identifying each single particle with an optimized combination of all
subdetector information:
charged tracks in the tracker and energy deposits in the ECAL and HCAL,
as well as signals in the the muon system and the preshower detector.
This procedure categorizes all particles into five types: muons, electrons,
photons, charged and neutral hadrons. The energy calibration is performed
separately for each particle type.

Electron candidates are reconstructed from clusters of energy deposited in the
ECAL. The clusters are first matched to track seeds in the pixel detector.
The trajectory of the electron candidate is reconstructed using a dedicated
modelling of the electron energy loss. Finally, the particle-flow algorithm
further distinguishes electrons from charged pions using a multivariate
approach~\cite{particleflow2}.

Muon candidates are identified by reconstruction algorithms
using signals in the silicon tracker and muon system.
The tracker muon algorithm
starts from tracks found in the tracker and then associates them with
matching signals in the muon chambers. The global muon algorithm starts
from standalone muons and then performs a global fit combining
signals in the tracker and muon system.

Jets from the fragmentation of quarks and gluons are reconstructed
from all particles found by the particle-flow algorithm using
the anti-$\kt$ jet clustering method~\cite{antikt} with the distance
parameter of $R = 0.5$, as implemented in \textsc{Fastjet} version
{2.4}~\cite{Cacciari:2005hq,fastjet1,fastjet2}. Small
corrections~\cite{Chatrchyan:2011ds} are applied as a function of $\eta$
and $\pt$ to the reconstructed jet energies.

A jet is identified as originating from a b quark using the combined
secondary-vertex (CSV) algorithm~\cite{CMS-PAS-BTV-11-003}, which provides
optimal b-tagging performance. The algorithm is based upon a likelihood test
that combines information about the impact parameter significance,
secondary-vertex reconstruction, and jet kinematics.
The small differences in the performance of the
$\cPqb$-tagging algorithm in data and MC simulation are accounted for by
data/MC scale factors. This is done by randomly removing
or adding \cPqb\ tags on a jet-by-jet basis, using the $\pt$- and $\eta$-dependent
scale factors discussed in~\cite{CMS-PAS-BTV-11-003}.

The missing transverse momentum in an event is defined as the negative vector
sum of the transverse momenta of all objects found from the particle-flow algorithm.

The vertex with the highest sum of $\pt^2$ of all associated tracks is taken
as the primary vertex of the hard collision.

\section{Event selection, signal and background estimation}
\label{sec:selection}

For this analysis we use an event selection similar to that adopted previously for
$\ttbar$ events in the lepton+jets channel~\cite{Chatrchyan:2011yy}. To reduce
the background from $\ttbar$ production, we apply higher jet $\pt$ thresholds.

Charged leptons from \PW-boson decays, which are themselves originating from
decays of heavy $\cPqt$-quark-like objects, are expected to be isolated from
nearby jets. A lepton isolation variable is calculated by summing the transverse
momenta of all reconstructed particles inside a cone defined as
$\Delta R=\sqrt{(\Delta \phi)^2+(\Delta \eta)^2}=0.3$, where
$\Delta\phi$ and $\Delta\eta$ are the azimuthal angle and pseudorapidity
differences with respect to the lepton direction. The lepton isolation variable
is equal to this sum divided by the lepton's $\pt$.

Events with exactly one isolated lepton and at least four jets with $|\eta|<2.4$
are selected. Jets that are within a cone of $\Delta R=0.3$ around the lepton
direction are not considered. At least one jet must be identified as originating
from a $\cPqb$ quark. The thresholds for the lepton $\pt$ are driven by
the trigger requirements described in Section~\ref{sec:data}. The lepton track
must have an impact parameter transverse to the beam direction with respect to
the primary vertex of less than 0.02\unit{cm} and along the beam direction of less than
1\unit{cm}. The missing $\pt$ in the event must be greater than 20\GeV.

The selection of the $\Pe$+jets events requires exactly one electron
with $\pt>35$\GeV and $|\eta|<2.5$, electron isolation $<0.1$,
and at least four jets with $\pt> 120$, 90, 50, and 35\GeV.
The selection for the $\Pgm$+jets channel requires exactly one muon with
$\pt>35$\GeV or $\pt>42$\GeV for two running periods with different trigger
conditions, $|\eta|<2.1$, muon isolation $< 0.125$, and at least four jets with
$\pt>120$, 90, 30, and 30\GeV. The thresholds for the two highest-$\pt$ jets are
selected to maximize the signal-to-background ratio. The thresholds for the lepton
$\pt$ and the third and fourth highest-$\pt$ jets are driven by the trigger
conditions.

Table~\ref{tab:events_1} lists the number of observed and expected events for
the various background sources after selection.
The expected numbers of background events are calculated from the cross
sections and integrated luminosities given in the table.
The cross section for $\ttbar$ production is taken from a previous CMS
measurement~\cite{Chatrchyan:2011yy}. All other cross sections are computed with
the \textsc{mcfm} program~\cite{MCFM}.
In the case of the e+jets channel, the small multijet background is
estimated from data by fitting the missing-$\pt$ distribution with shapes predicted
by the MC simulation.
The uncertainties shown include systematic uncertainties in the efficiency and
acceptance. Uncertainties are strongly correlated for all sources. Uncertainties
in the cross sections and the integrated luminosity are not included.

The fraction of $\ttbar$ events retained by our selection is 0.7\% for the $\Pgm$+jets
channel and 0.5\% for the $\Pe$+jets channel.

\begin{table*}[htbp]
\topcaption{
Background cross sections, number of events observed and background events
predicted for the $\Pe$+jets and $\Pgm$+jets samples. The predicted
numbers of events are normalized to the integrated luminosity
(except for the multijet events in the $\Pe$+jets channel, see text).
}
\label{tab:events_1}
\begin{center}
\begin{tabular}{lccc}
\hline\hline
& & $\Pe$+jets & $\Pgm$+jets \\
\hline
Integrated luminosity           &     & 4.98\fbinv & 4.90\fbinv \\
\hline
Background process              & Cross section & Events         & Events \\
\hline
$\ttbar$             & 154\unit{pb}       & ~$3950\pm490$ & ~$5460\pm670$ \\
\PW+jets                          & ~31\unit{nb}   & ~~$462\pm55$  & ~~$750\pm110$~ \\
Single-$\cPqt$ production  & ~85\unit{pb}       & ~~$208\pm24$~ & ~~$336\pm45$~~ \\
\cPZ+jets, $\PW\PW$, $\PW\cPZ$, $\cPZ\cPZ$              & ~3.1\unit{nb} & ~~~$49\pm8$~  & ~~~$69\pm11$ \\
Multijets                       &              & ~~~$78\pm9$~  & ~~~~$5\pm5$ \\
Total background                &              & ~$4750\pm560$ & ~$6620\pm800$ \\
\hline
Total observed                  &              & ~4734~~~~      & ~6448~~~~ \\
\hline\hline
\end{tabular}
\end{center}
\end{table*}

The comparisons between the data and the simulated background of multiple
distributions for the final objects (leptons, jets, and missing transverse
momentum) and their combinations have been performed, both for the initial
$\ttbar$ selection~\cite{Chatrchyan:2011yy} and for the final $\PQtpr$
requirements. In all cases, the data are in agreement with the background model
predictions, within the statistical uncertainties.

Table~\ref{tab:signal} shows the theoretical cross sections for the signal process
for various $\PQtpr$-quark masses, along with the efficiencies of the event
selection for the $\Pe$+jets and $\Pgm$+jets channels and the expected numbers
of signal events. The $\PQtpr\PAQtpr$
production cross sections are computed using \textsc{hathor}~\cite{HATHOR}. The
efficiencies include the branching fraction of the $\PQtpr\PAQtpr$ system
into a single-lepton final state, which can be obtained from the branching fractions
for $\PW\to\ell\Pgn$ and $\PW\to\cPq\cPaq'$.
The uncertainties quoted are the statistical uncertainties from the MC simulation.

\begin{table*}[htbp]
\topcaption{Theoretical cross sections~\cite{HATHOR}, selection efficiencies, and
numbers of expected events for the $\PQtpr\PAQtpr$ signal with different
$\PQtpr$ masses in the $\Pe$+jets and $\Pgm$+jets channels. The
efficiencies include the branching fraction of the $\PQtpr\PAQtpr$
system into a single-lepton final state.}
\label{tab:signal}
\begin{center}
\begin{tabular}{cccccc}
\hline\hline
$M_{\PQtpr}$ (\GeVns) & Cross section (pb) & $\Pe$+jets eff. (\%) & Events & $\Pgm$+jets eff.  (\%) & Events \\
\hline
400 & 1.41   & $4.3\pm0.1$ & 302 & $5.4\pm0.1$ & 373 \\
425 & 0.96   & $4.4\pm0.1$ & 210 & $5.6\pm0.1$ & 263 \\
450 & 0.66   & $4.7\pm0.1$ & 155 & $6.0\pm0.1$ & 194 \\
475 & 0.46   & $4.7\pm0.1$ & 108 & $6.1\pm0.1$ & 137 \\
500 & 0.33   & $4.8\pm0.1$ &  79 & $6.2\pm0.1$ & 100 \\
525 & 0.24   & $4.7\pm0.1$ &  56 & $6.4\pm0.1$ &  75 \\
550 & 0.17   & $4.9\pm0.1$ &  41 & $6.5\pm0.1$ &  54 \\
575 & 0.13   & $4.7\pm0.1$ &  30 & $6.6\pm0.1$ &  42 \\
600 & 0.092  & $4.7\pm0.1$ &  22 & $6.6\pm0.1$ &  30 \\
625 & 0.069  & $4.8\pm0.1$ &  16 & $6.5\pm0.1$ &  22 \\
\hline\hline
\end{tabular}
\end{center}
\end{table*}

\section{Mass reconstruction}

We perform a kinematic fit of each event to the
$\PQtpr\PAQtpr\to\PW\cPqb\PW\cPaqb\to\ell\Pgn\cPqb\cPq\cPaq'\cPaqb$
process. There are two steps in the reconstruction of the $\PQtpr$-quark mass: the
assignment of reconstructed objects to the quarks, and a kinematic fit to improve
the resolution of the reconstructed mass of the $\PQtpr$-quark candidates.
The four-momenta resulting from the kinematic fit of the particles in the final state
must satisfy the following three constraints, where $m$ is the invariant mass of
the corresponding particles in parentheses,
$M_\PW$ is the $\PW$-boson mass, $\Mfit$ is a free parameter in the fit
(reconstructed $\PQtpr$ mass), and $\ell$ stands for electron or muon:
\begin{eqnarray}
m(\ell\Pgn)= M_\PW, \\
m(\cPq\cPaq')= M_\PW, \\
m(\ell \Pgn \cPqb) = m(\cPq\cPaq'\cPqb) = \Mfit.
\end{eqnarray}
Here $\ell$, $\Pgn$, $\cPqb$ denote either particle or antiparticle.

The reconstructed objects in the event are the charged lepton, the neutrino,
and four or more jets. For the neutrino, only its transverse momentum
can be measured as the missing transverse momentum in the event.
The $z$ component of the neutrino momentum can be determined with two solutions
from the kinematic constraints.
The four quarks in the final state manifest themselves as jets and their momenta
are measured. Thus, all but one of the  momentum components of the considered final
system are measured. With one unknown and three constraints, a kinematic fit is
performed by minimizing the $\chi^2$ computed from the difference between the
measured momentum of each reconstructed object and its fitted value, divided by
its uncertainty.

We have studied different strategies for pairing the observed jets with the four
quarks from the decay of the $\PQtpr$ and $\PAQtpr$ quarks to find
the best separation between the $\PQtpr\PAQtpr$ signal and
the $\ttbar$ background. In events with exactly four jets,
we consider all possible jet-quark assignments.
To reduce the number of combinations, we choose only those in which at least one
b-tagged jet is assigned to a \cPqb\ quark from the $\PQtpr\PAQtpr$ decay.
In events with more than four jets, we take the five jets having the highest
$\pt$ values, and consider all combinations of four out of these five jets.
In each event, the kinematic fit is carried out for each jet-quark assignment,
and the jet-quark assignment with the smallest $\chi^2$ value is chosen.
This procedure selects the correct jet-quark assignment in 36--40\% of
the simulated $\PQtpr\PAQtpr$ events over a $\PQtpr$-quark
mass range of 400--625\GeV for all jet multiplicities together. For
$\ttbar$ events this fraction is much lower, about 19\%,
because the jets from the decays of the $\cPqt$ and $\cPaqt$
quarks are softer than
from $\PQtpr$ and $\PAQtpr$ decays and, therefore, are
less likely to be among the five highest-$\pt$ jets in the event.
The $\chi^2$ value does not distinguish the $\PQtpr\PAQtpr$ signal
from the $\ttbar$ background because both processes satisfy the
fit hypothesis, but using the smallest value for each event does increase
the fraction of correct quark assignments. Since a restriction on $\chi^2$
does not improve the signal-to-background ratio, no such restriction is applied.

\begin{figure*}[htp]
\begin{center}
\includegraphics[width=0.8\textwidth]{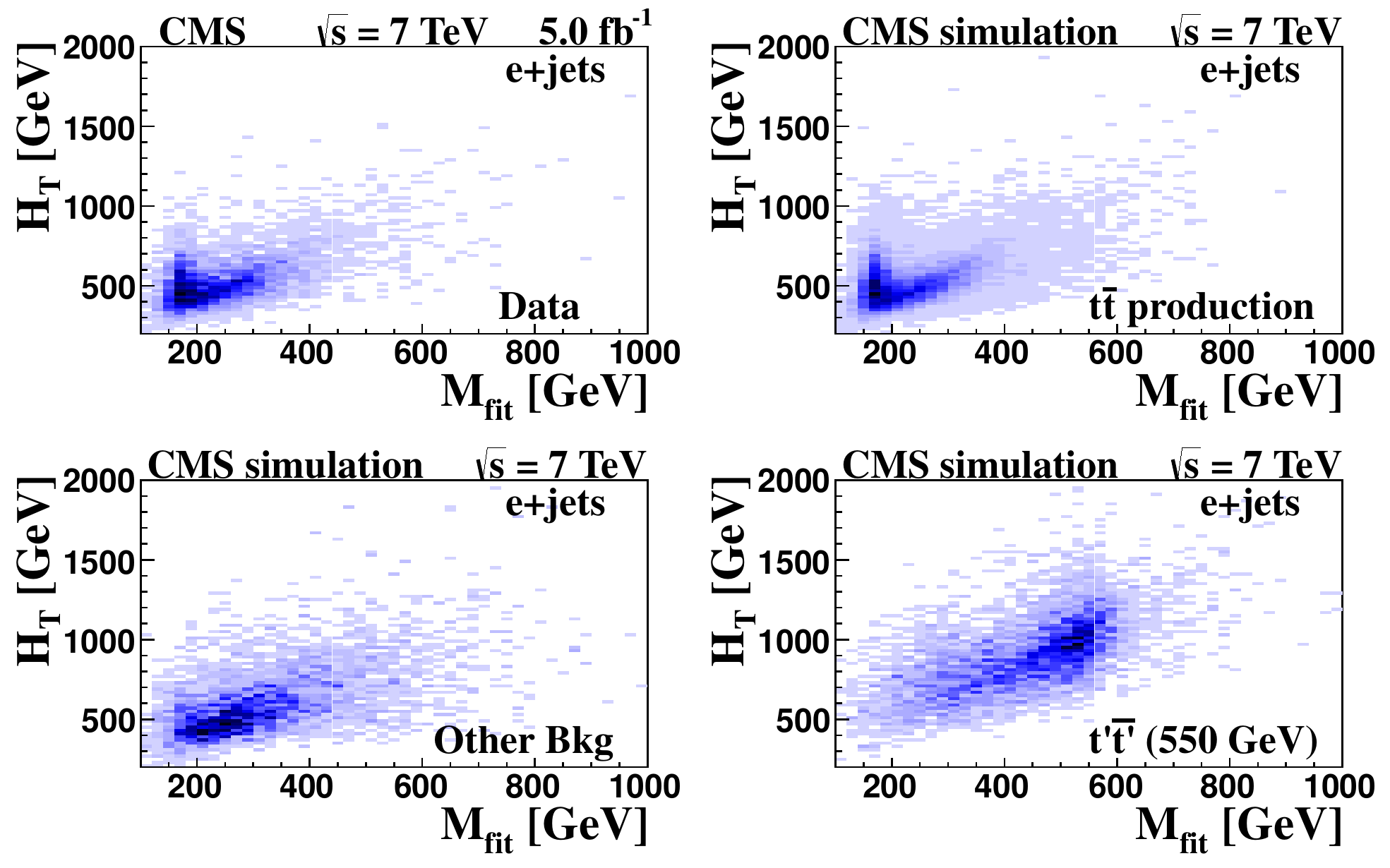}
\caption{$\HT$ versus $\Mfit$ for the $\Pe$+jets channel from data
(top left), and simulations of $\ttbar$ production (top right), other backgrounds
(bottom left), and $\PQtpr\PAQtpr$ production (bottom right) for
$M_{\PQtpr}=550$\GeV.}
\label{fig:ejets_templates}
\end{center}

\begin{center}
\includegraphics[width=0.8\textwidth]{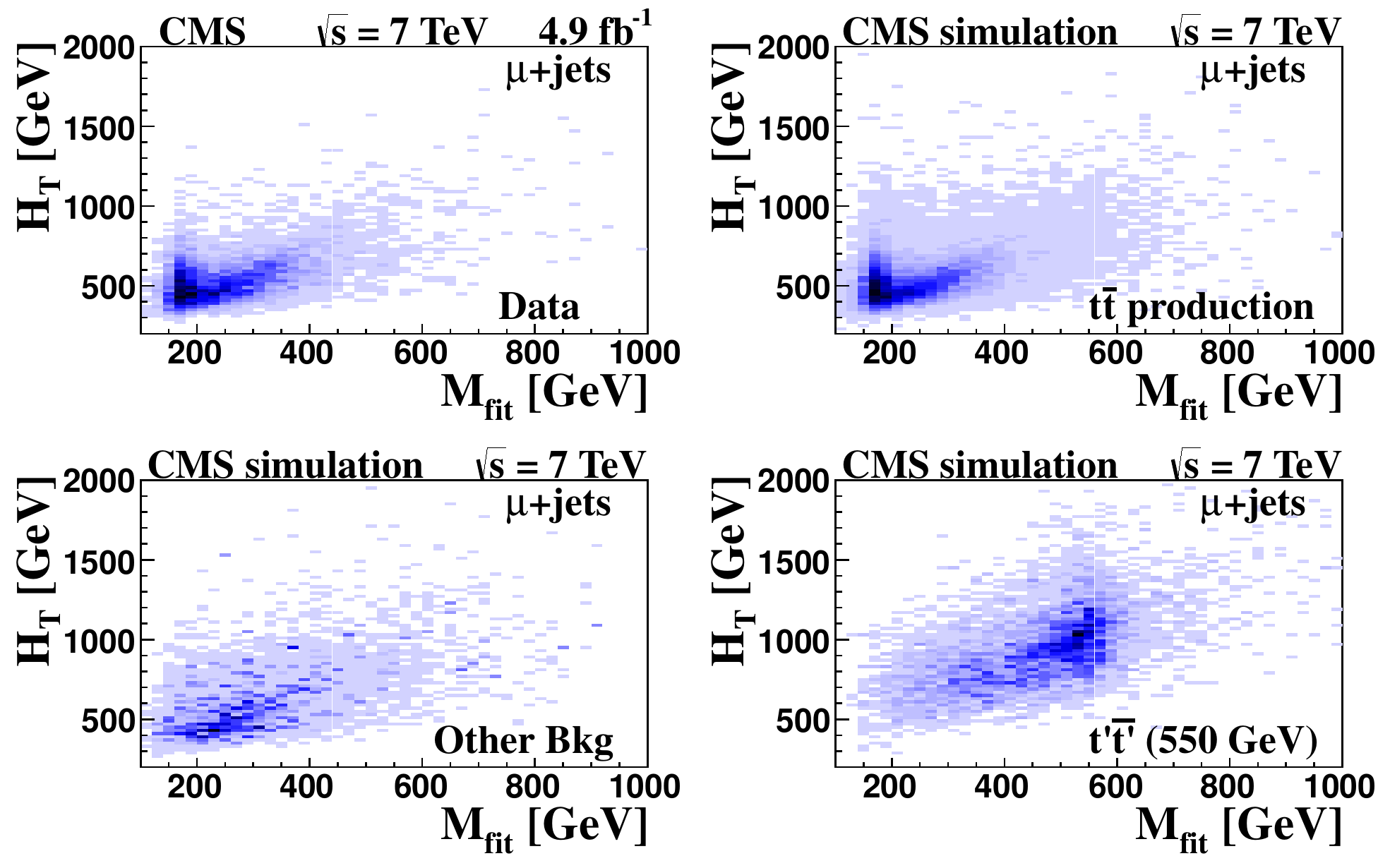}
\caption{$\HT$ versus $\Mfit$ for the $\Pgm$+jets channel from data
(top left), and simulations of $\ttbar$ production (top right), other backgrounds
(bottom left), and $\PQtpr\PAQtpr$ production (bottom right) for
$M_{\PQtpr}=550$\GeV.}
\label{fig:mjets_templates}
\end{center}
\end{figure*}

\begin{figure*}[htbp]
\begin{center}
\includegraphics[width=0.45\textwidth]{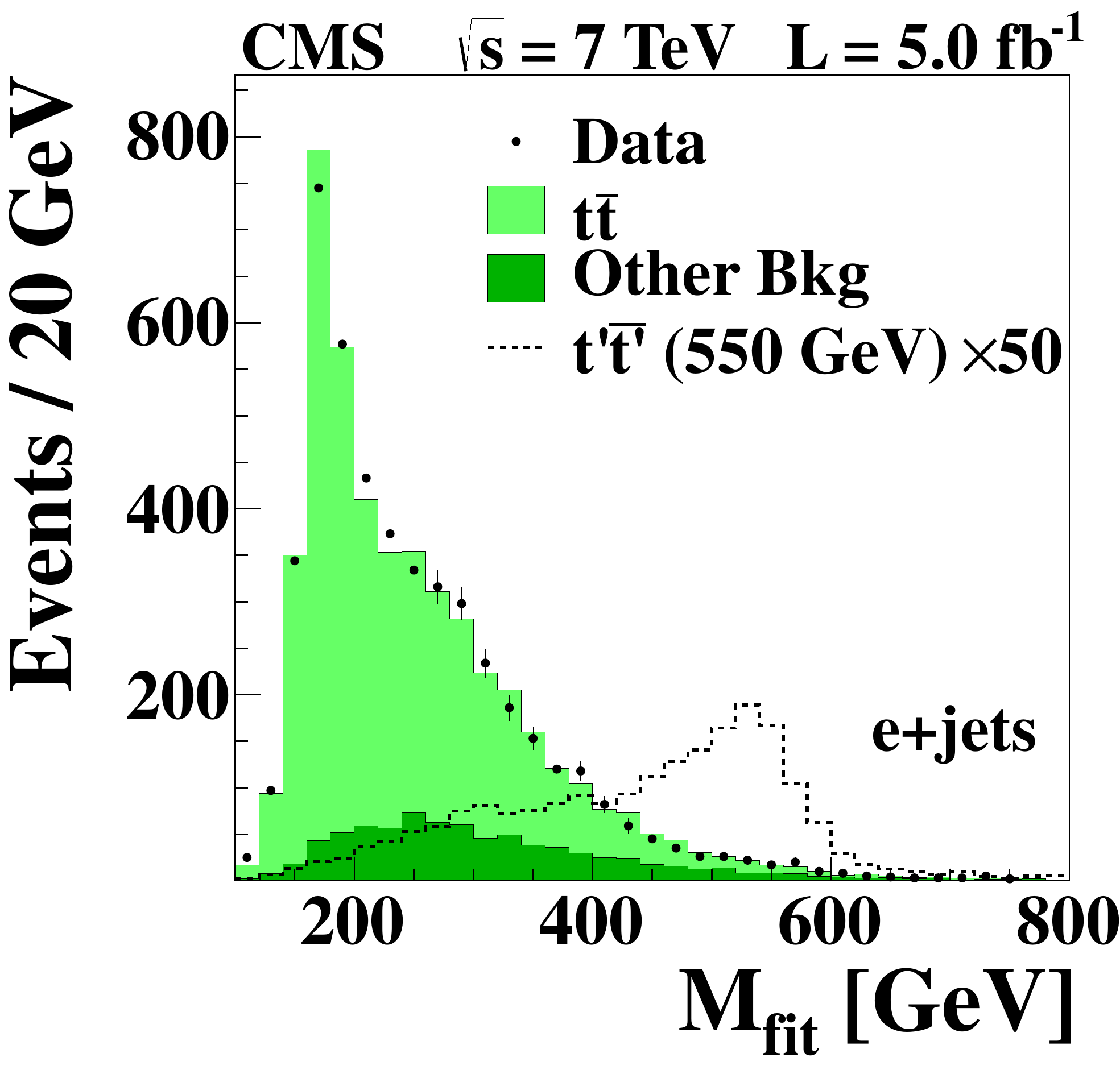}
\includegraphics[width=0.45\textwidth]{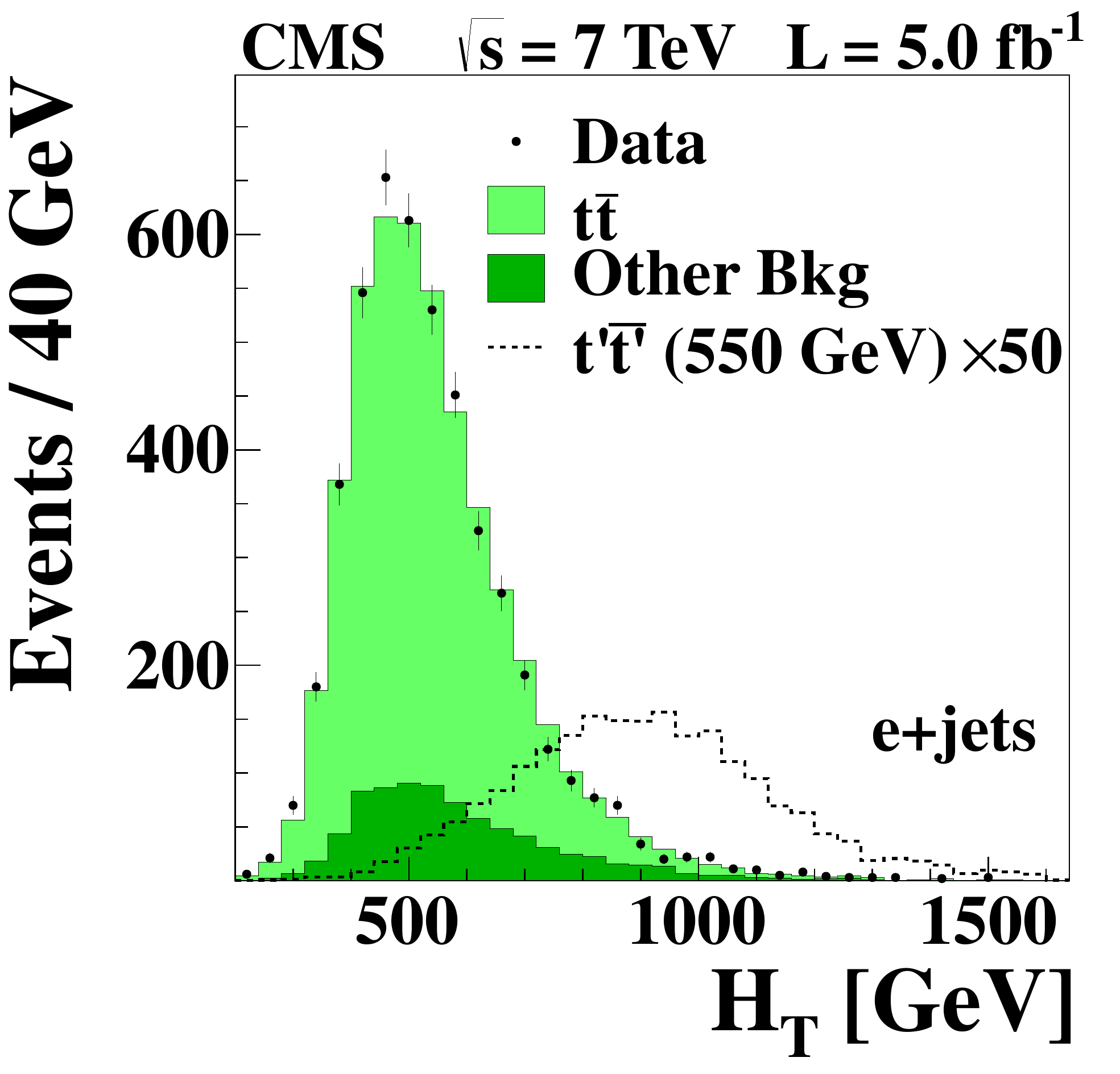} \\
\includegraphics[width=0.45\textwidth]{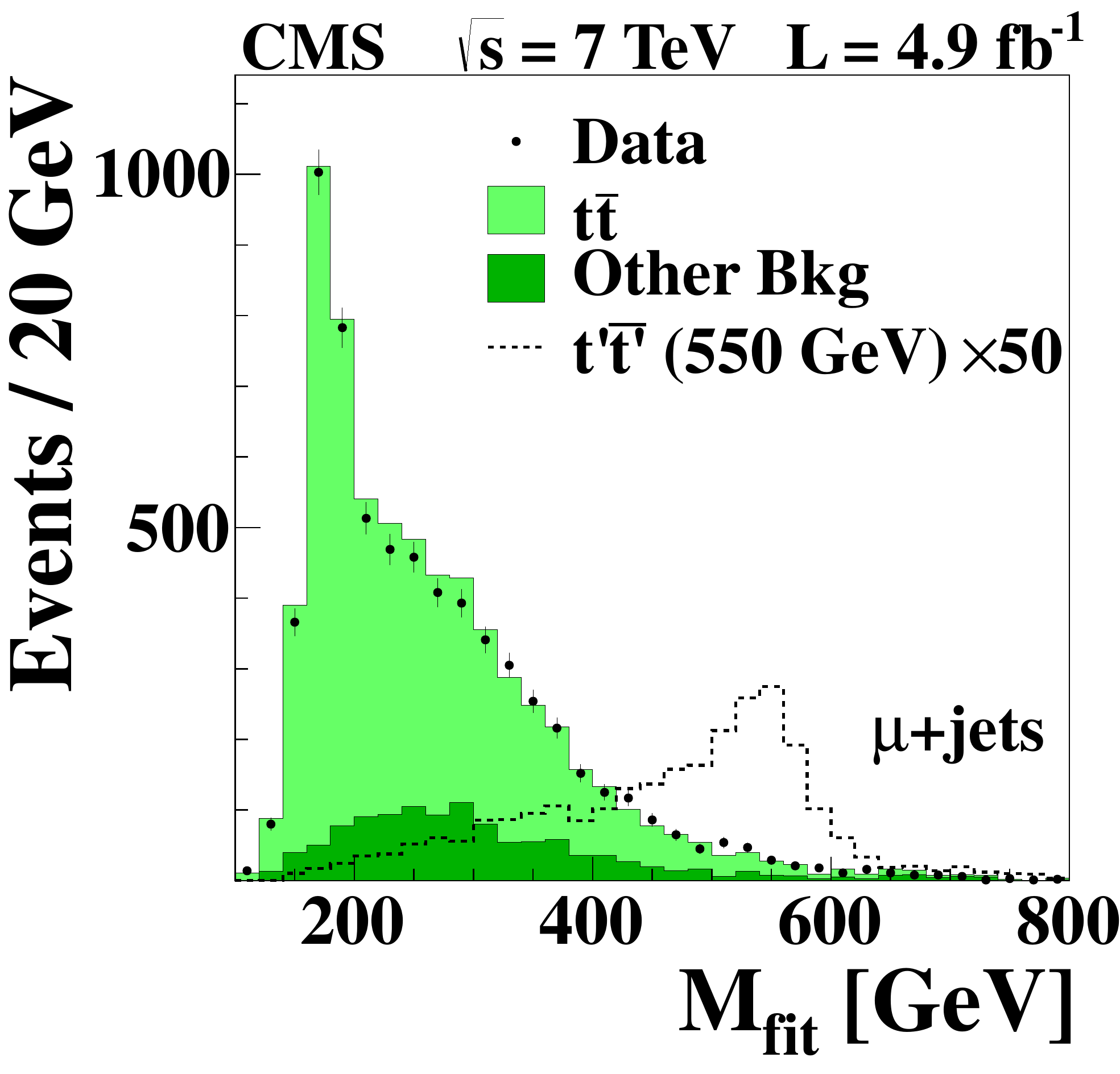}
\includegraphics[width=0.45\textwidth]{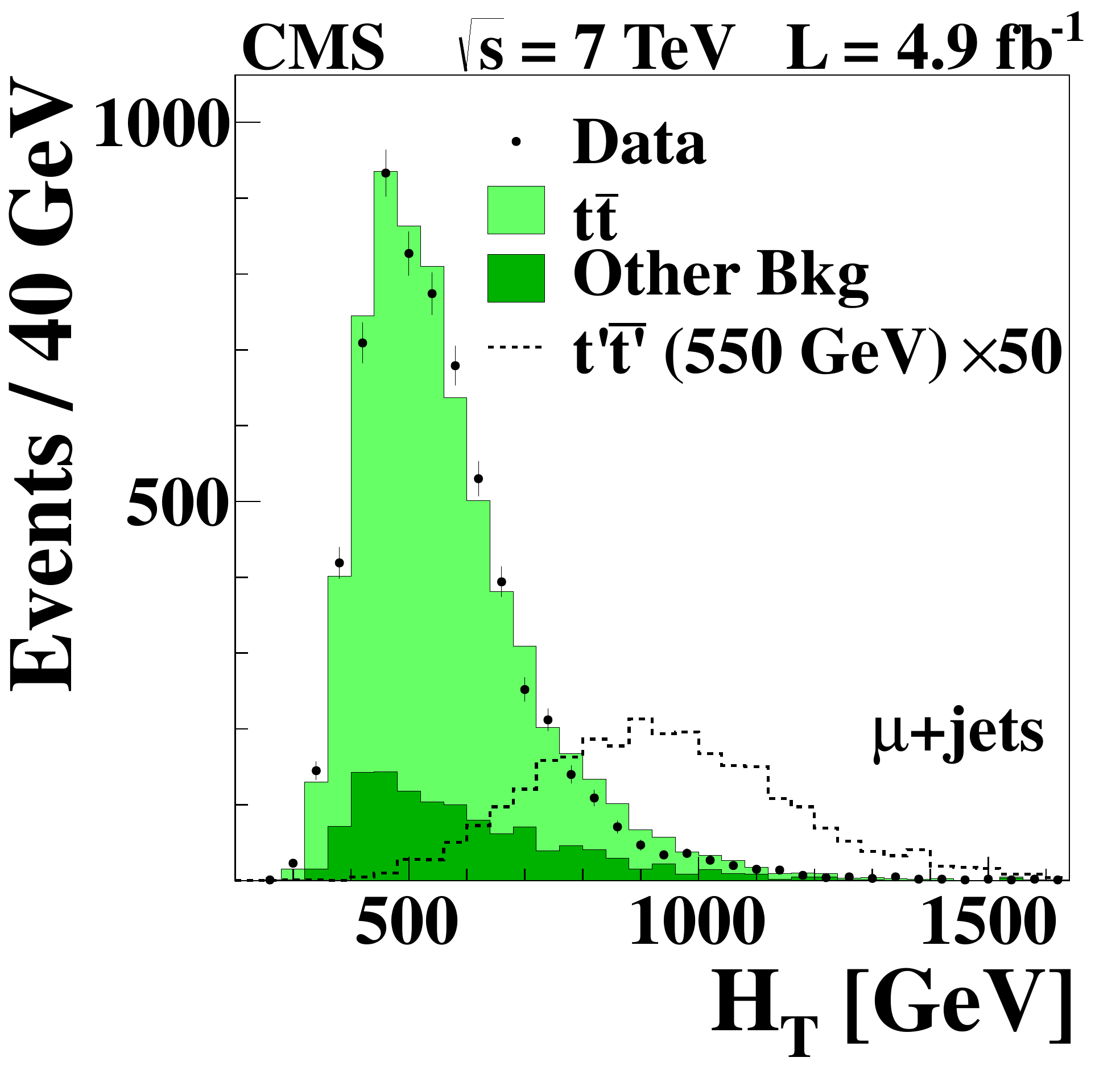}
\caption{Distributions of $\Mfit$ (left) and $\HT$ (right) for the
$\Pe$+jets (top) and $\Pgm$+jets (bottom) channels. The data are shown
as points, the simulated backgrounds as shaded histograms, and the expected
signal for a $\PQtpr$ mass of 550\GeV
as dashed histograms (multiplied by a factor of 50 to improve visibility).}
\label{fig:mfit_ht}
\end{center}
\end{figure*}

Figures~\ref{fig:ejets_templates} and \ref{fig:mjets_templates} show the
two-dimensional $\HT$ versus $\Mfit$ distributions for the data,
$\ttbar$ simulation, the other simulated backgrounds, and the
$\PQtpr\PAQtpr$ simulation with a particular $\PQtpr$ mass of 550\GeV
in the $\Pe$+jets and $\Pgm$+jets channels, respectively.
Figure~\ref{fig:mfit_ht} shows the corresponding $\Mfit$ and $\HT$
projections. The integrated luminosities given in Table~\ref{tab:events_1} are
used for the normalization of the background processes. The data are found to
be in agreement with the predicted background $\Mfit$ and $\HT$
distributions.
The $\ttbar$ events that pass the selection criteria either have high-$\pt$
$\cPqt$ and $\cPaqt$ quarks that produce high-$\pt$ jets in
their decays or they have high-$\pt$ jets from initial-state gluon radiation.
The former class of events is responsible for the relatively narrow peak in the
$\Mfit$ distribution at the \cPqt-quark mass. The $\Mfit$
distribution of the latter class of events is broad and typically
populates the region above the \cPqt-quark mass, leading to the observed
high-mass tail in the $\Mfit$ distribution.

\section{Computation of \texorpdfstring{$\PQtpr\PAQtpr$}{t' t'-bar} cross section limits}

The two-dimensional distributions of $\HT$ versus $\Mfit$,
such as those shown in Figs.~\ref{fig:ejets_templates} and
\ref{fig:mjets_templates}, are used to search for a $\PQtpr\PAQtpr$
signal in the data. Simulated $\PQtpr\PAQtpr$ signal distributions
are produced for $\PQtpr$ masses from 400 to 625\GeV in 25\GeV steps.
We do not use the two-dimensional histograms directly because
it is not possible to simulate enough events to adequately populate all
bins of the distributions for both signal and background. Therefore, we
employ a new procedure that combines bins.

All the background distributions are added together to obtain the expected
background event yield in each bin of the $\HT$ versus $\Mfit$
histogram. Then the projections of the two-dimensional signal and background
histograms onto the $\HT$ and $\Mfit$ axes are separately fitted with
analytic functions. Next, we compute the expected signal-to-background ($s/b$)
ratio for each two-dimensional bin as the product of the values
of the two one-dimensional-bin fit functions for the signal and for the
background at the bin center. This procedure of fitting the projections and neglecting
their correlations is chosen because it reduces the sensitivity of $s/b$
ordering to statistical fluctuations in the simulated samples. These
functions are used only to define the ordering of the bins. All
two-dimensional bins are then sorted in increasing order of the expected
$s/b$ ratio, which we call the $s/b$ rank.

\begin{figure}[htb]
\begin{center}
\includegraphics[width=0.45\textwidth]{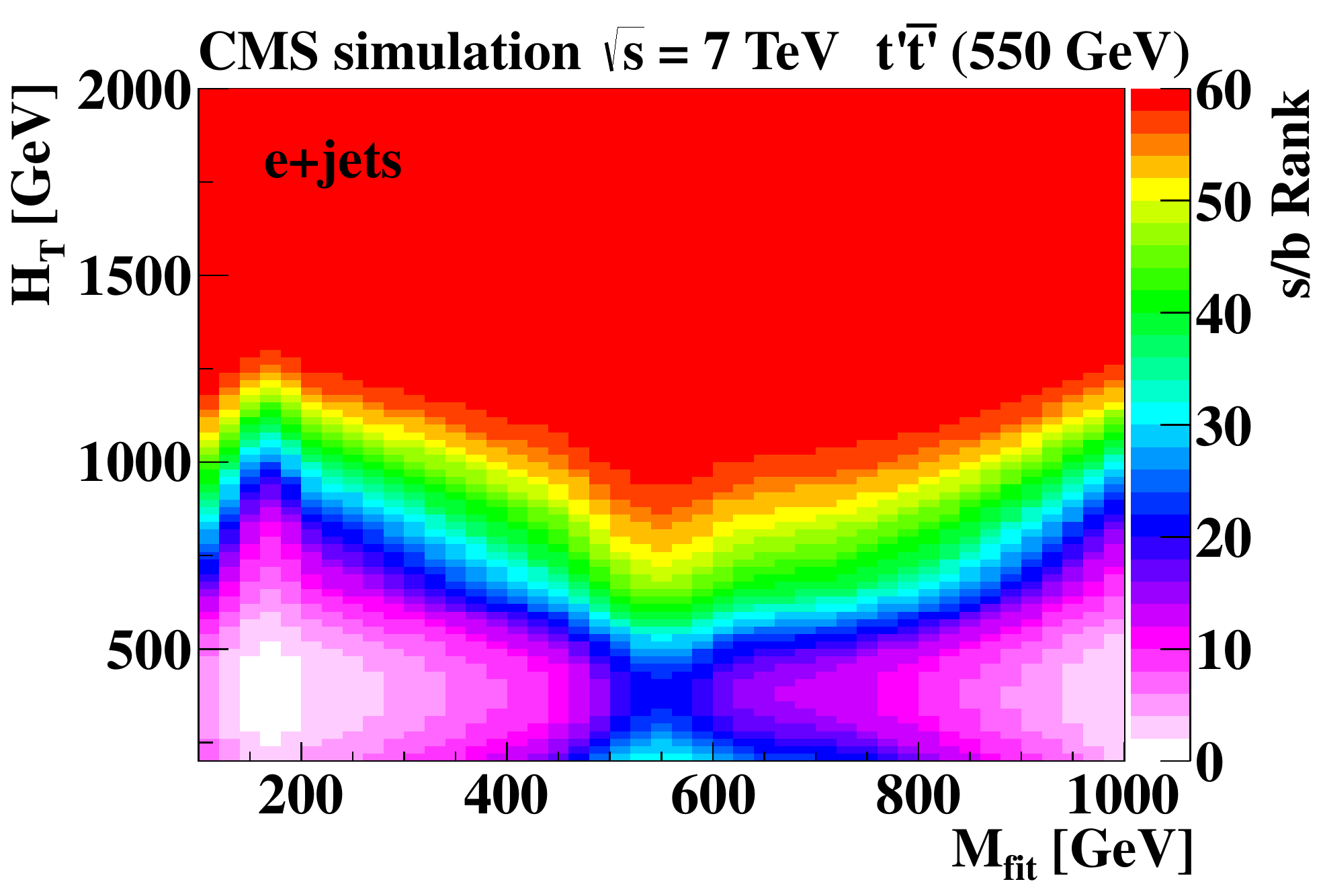}
\includegraphics[width=0.45\textwidth]{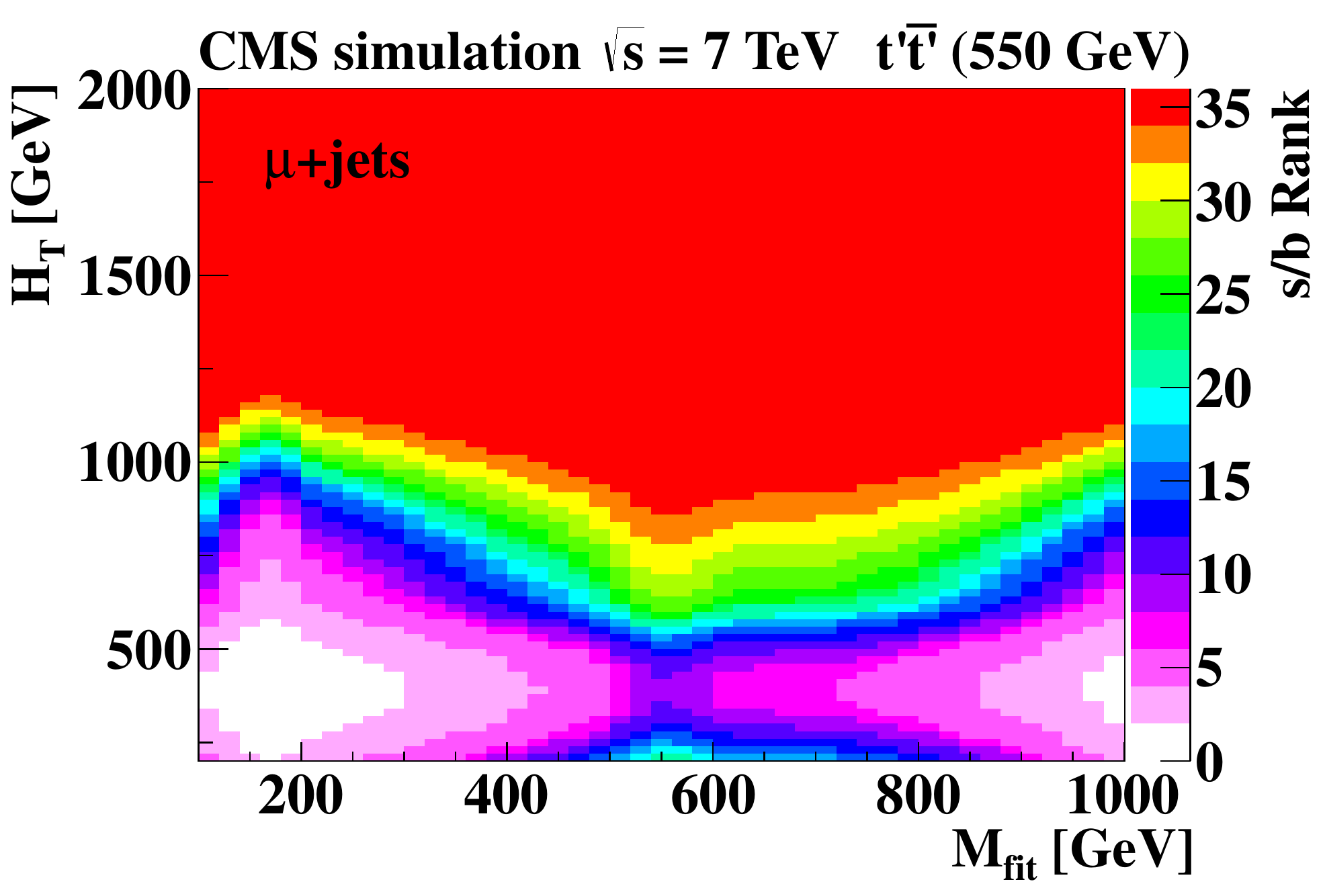}
\caption{Map of the merged bins in the $\HT$ versus $\Mfit$
plane for a $\PQtpr$ quark with a mass of 550\GeV for the $\Pe$+jets
(\cmsLeft) and $\Pgm$+jets (\cmsRight) channels. The colour represents bins merged
according to increasing signal-to-background ($s/b$) ratio.
The vertical colour axis is labelled by $s/b$ rank and corresponds to the bin
index of the one-dimensional histograms of Fig.~\ref{fig:templates}.
}
\label{fig:maps}
\end{center}
\end{figure}

We then merge the two-dimensional bins that are adjacent after ordering by $s/b$
ratio so that the fractional statistical uncertainty of both the signal and
the background predictions is below 20\% in all bins.
We select the 20\% value as a compromise between two effects. Increasing this
value would increase the $\PQtpr$ signal sensitivity, but would also
increase the potential biases in the $\PQtpr\PAQtpr$ cross section
measurement, as determined from MC-generated ``pseudo-experiments", described
below.
Figure~\ref{fig:maps} shows the colour-coded maps of the merged bins obtained for
the simulation of a $\PQtpr$ quark with a mass of 550\GeV. The colour
represents the rank of the bin in the $s/b$ ordering. A higher rank corresponds
to a higher $s/b$ value.

\begin{figure}[htb]
\begin{center}
\includegraphics[width=0.45\textwidth]{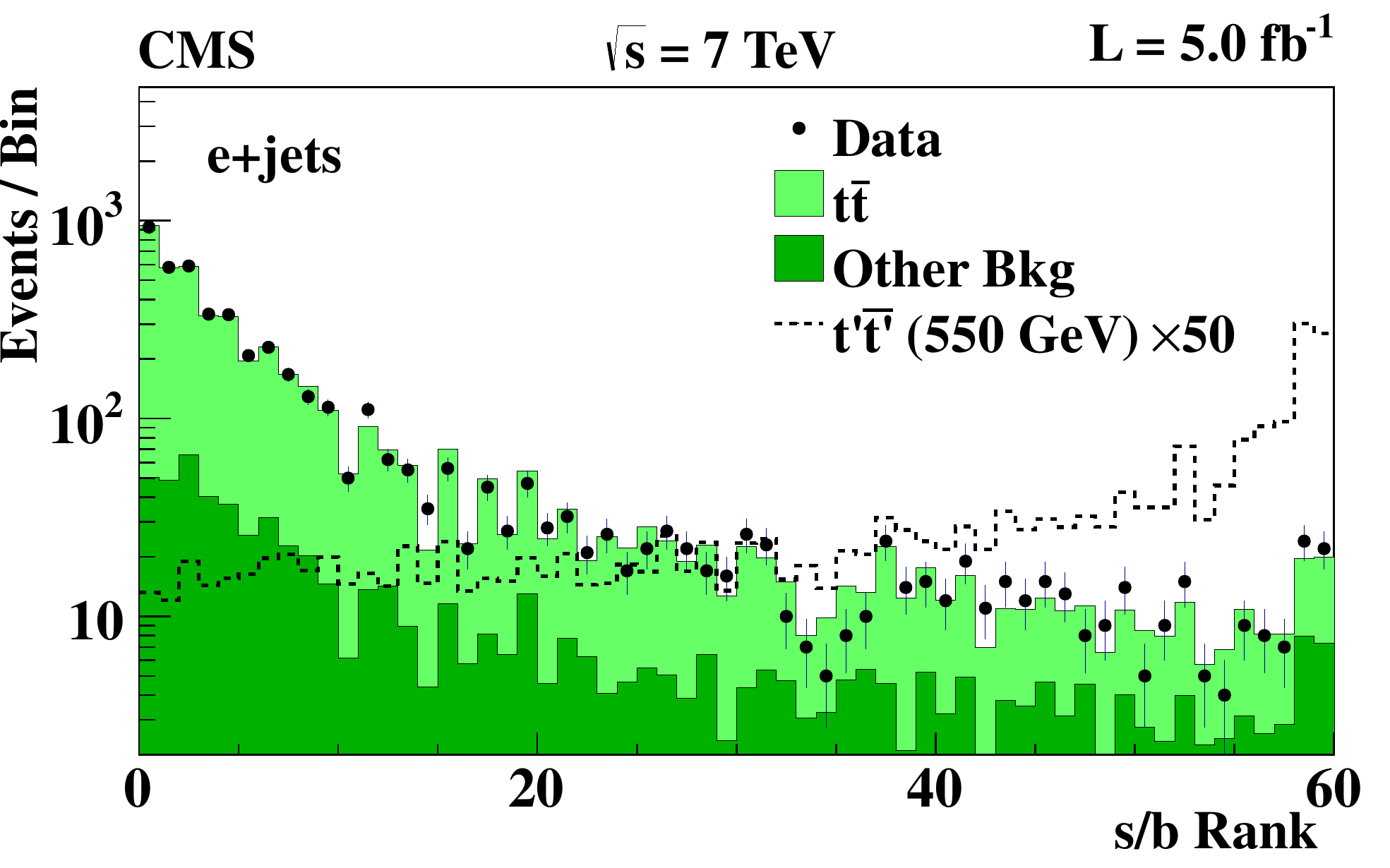}
\includegraphics[width=0.45\textwidth]{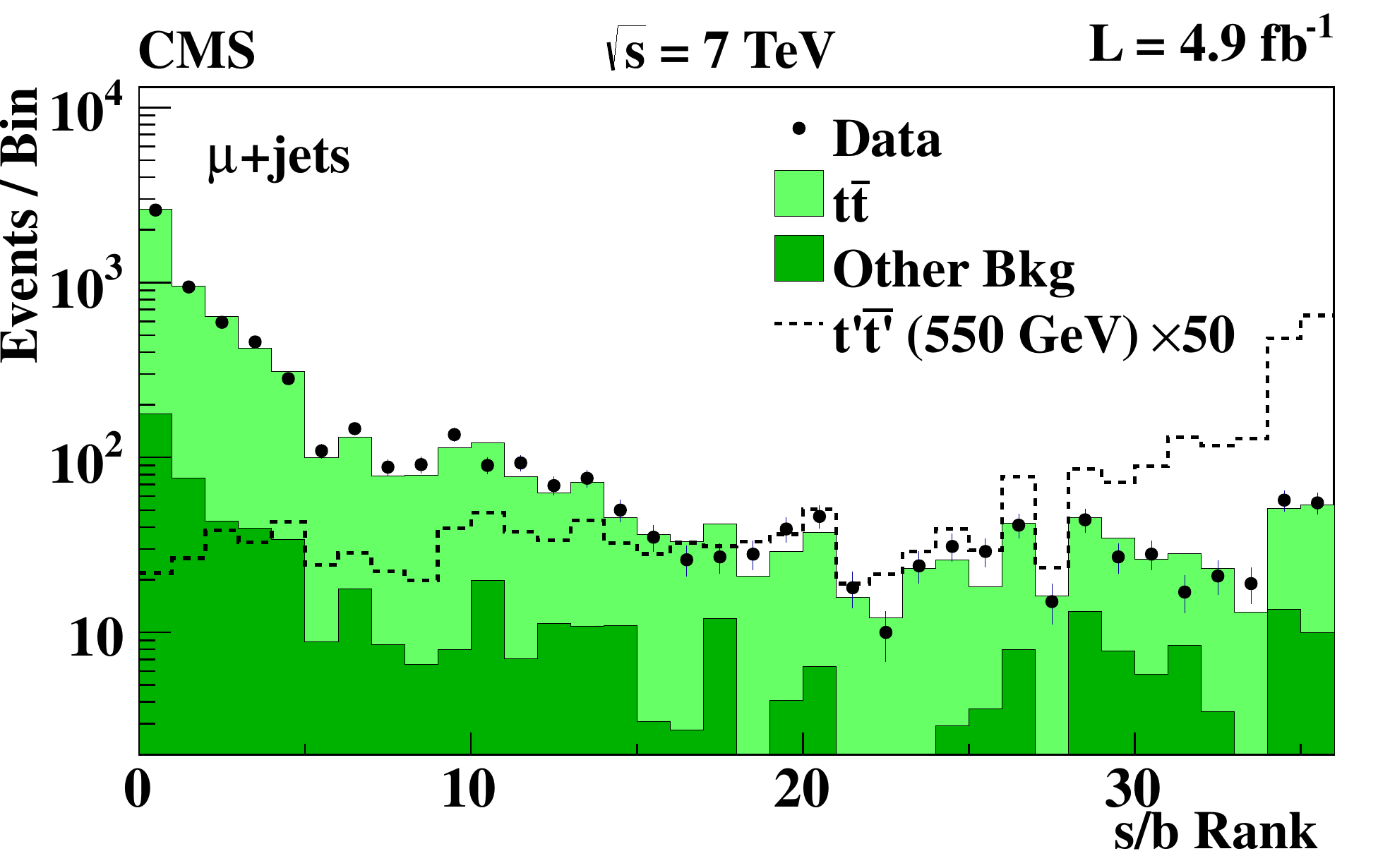}
\caption{Number of events per bin in the two-dimensional $\HT$ versus
$\Mfit$ histogram after bin merging, as a function of the
signal-to-background ($s/b$) rank for the $\Pe$+jets (\cmsLeft) and $\Pgm$+jets
(\cmsRight) channels. The data are shown by the points, the simulated
$\ttbar$ and other background distributions by the histograms,
and the prediction for a $\PQtpr\PAQtpr$ signal with a $\PQtpr$
mass of 550\GeV by the dotted lines (multiplied by a factor of 50 for easier
viewing).
}
\label{fig:templates}
\end{center}
\end{figure}

In Fig.~\ref{fig:templates}, the number of events in the merged bins is plotted
versus $s/b$ rank. In these histograms, signal events will predominantly cluster
towards the right, and background events towards the left. These one-dimensional
histograms are used as input to the $\PQtpr\PAQtpr$ cross section computation,
and we will refer to these distributions as templates in the following.
The data agree with the predicted background distributions in Fig.~\ref{fig:templates},
with no evidence for a $\PQtpr$ signal. Thus, we use the results to set upper
limits on the $\PQtpr\PAQtpr$ cross section as a function of $\PQtpr$
mass.

The computation of the limits for the $\PQtpr\PAQtpr$ cross section
uses the \CL{s} criterion~\cite{CLs1,CLs2}. The first step is to perform
a likelihood fit to the data. We group the background in two components:
the larger one due to $\ttbar$ production and the smaller one from all EW
processes ($\PW$+jets, $\cPZ$+jets, single-$\cPqt$,
and diboson production) and from multijet processes. Each background
component is normalized to its expected yield and multiplied with a scale
factor that is a free parameter in the fit.
The $\PQtpr\PAQtpr$ cross section, $\sigma$, is also a free parameter
in the fit. The following likelihood ratio is used as the test statistic:

\begin{equation}
t(q|\sigma) = \begin{cases}
L(q|\sigma,\hat\alpha_\sigma)/L(q|\hat\sigma, \hat\alpha) &\text{if } \sigma>\hat\sigma \\
1 & \text{if } \sigma\leq\hat\sigma. \end{cases}
\end{equation}

Here, $L(q|\sigma, \alpha)$ is the likelihood of the data having the value $q$ for
the parameter of interest and the nuisance parameters $\alpha$. The nuisance parameters
account for effects that give rise to systematic uncertainties in the templates and include
the normalizations of the background components. We do not include the per-bin statistical
uncertainties on the signal and background predictions in the likelihood fit because their
effects were found to be negligible after applying the bin-merging procedure described
above.
The likelihood reaches its maximum
when $\sigma=\hat\sigma$ and $\alpha=\hat\alpha$. The symbol $\hat\alpha_\sigma$
refers to the values of the nuisance parameters $\alpha$ that maximize the conditional
likelihood at a given value of $\sigma$.

Using the asymptotic approximation for the test statistic described
in~\cite{asymptoticCLs}, the probability to observe a value $t$ for the
likelihood ratio that is larger than the observed value $t_\text{obs}$ is determined.
This is done by producing samples of pseudo-experiments in which the expected
numbers of signal and background events are allowed to vary according to their
statistical and systematic uncertainties. For the pseudo-experiments generated
with background only, this probability is denoted by \CL{b}. For pseudo-experiments
with a cross section $\sigma$ for the $\PQtpr\PAQtpr$ signal, this probability
is denoted by $\CL{s+b}(\sigma)$, which is a function of $\sigma$.
The upper limit at the 95\% confidence level (\CL{}) for the $\PQtpr\PAQtpr$
cross section is the value of $\sigma$ for which $\CL{s} = \CL{s+b}/\CL{b} = 0.05$.
To determine the limits for both lepton channels combined, we simultaneously
fit the histograms from both channels, accounting for correlations among
the nuisance parameters, and then apply the $\CL{s}$ method described above.

\section{Systematic uncertainties}
\label{sec:syst}

The signal and background predictions are subject to systematic uncertainties.
Below, we describe all sources of systematic uncertainties that have been considered.
They can be divided into two categories: uncertainties that only impact the
normalization of the signal and background templates, and uncertainties that also
affect the shapes of the distributions.

The uncertainties in the $\ttbar$ cross section, electroweak and multijet
background normalizations, integrated luminosity, lepton efficiencies, and
data/MC scale factors affect only the normalization.

The uncertainty on the cross section for $\ttbar$ production is taken from the
CMS measurement of $154\pm18$\unit{pb} at $\sqrt{s} = 7$\TeV~\cite{Chatrchyan:2011yy}.
The predicted yields of the EW and multijet backgrounds are determined as
described in Section~\ref{sec:selection}. A 50\% uncertainty is assigned to the sum of these
two backgrounds in the likelihood fit to the data in order to account for
the uncertainty in the acceptance and the W+jets normalization.

The integrated luminosity affects the normalization of the $\PQtpr\PAQtpr$
signal and the background templates in a correlated way. The integrated
luminosity is known to a precision of 2.2\%~\cite{luminosity}.

Trigger efficiencies, lepton identification efficiencies, and data/MC scale
factors are obtained from data using decays of \cPZ\ bosons to dileptons.
Their uncertainties are included in the selection efficiency uncertainty.
They amount to 2\% for the $\Pgm$+jets channel and 3\% for the e+jets channel.

Uncertainties that affect the shapes of the distributions include those on
the jet energy scale, jet energy resolution, missing-$\pt$ resolution,
$\mathrm{b}$-tagging efficiency, number of multiple $\Pp\Pp$ interactions,
factorization/renormalization scale $Q$, matrix-element/parton-shower matching
threshold~\cite{matching}, and initial- and final-state radiation. To model
these uncertainties, we produce additional templates by varying the nuisance
parameter that characterizes the systematic effect by
$\pm$1 standard deviation. To determine the signal and background templates
used in the fit for any value of the nuisance parameter, we interpolate the
content of each bin between the varied and nominal templates. This procedure
is often referred to as vertical morphing.

The energy of all jets is obtained using the standard CMS jet energy
calibration constants~\cite{Chatrchyan:2011ds}. The sum of the four-momenta
of the jets is 100\% correlated with the measured missing
$\pt$. The jet energy scale uncertainty affects the normalization and
the shape of the $\HT$ vs.~$\Mfit$ distribution. This is taken into
account by generating $\HT$ vs.~$\Mfit$ distributions for values
of the jet energy scaled by $\pm1$ standard deviation of the $\eta$-
and $\pt$-dependent uncertainties from~\cite{Chatrchyan:2011ds}.

The energy resolution of jets in the simulation is better than in the data.
Therefore, random noise is added to the jet energies in the simulation to worsen
the resolution by 10\%, to match the actual resolution of the detector. To estimate
the corresponding uncertainty, the analysis is performed without smearing and with
20\% smearing.  The missing-$\pt$ resolution is also simultaneously
corrected for this effect.

The systematic uncertainty from the $\mathrm{b}$-tagging efficiency is
estimated by varying this efficiency by $\pm$1 standard deviation taken
from~\cite{CMS-PAS-BTV-11-003}.

To evaluate the uncertainties related to the modelling of multiple
interactions in the same beam crossing, the average number of interactions
in the simulation is varied by $\pm$8\% relative to the nominal value.

The uncertainty in the factorization/normalization scale $Q$, used for
the strong coupling constant ${\alpha}_s(Q^2)$, is estimated by using
two sets of simulated $\ttbar$ samples in which the $Q$ value is increased and
decreased by factors of two relative to the nominal value.

The uncertainty arising from the threshold for matching between matrix
elements and parton showers~\cite{matching} is estimated using two simulated
$\ttbar$ samples generated with the matching threshold varied up and down by
a factor of two from the default value.

The impact of initial- and final-state radiation is estimated using
a $\ttbar$ MC sample generated with {\POWHEG}, instead of
\textsc{MadEvent}/{\MADGRAPH}.

\begin{figure}[htbp]
\begin{center}
\includegraphics[width=\cmsFigWidth]{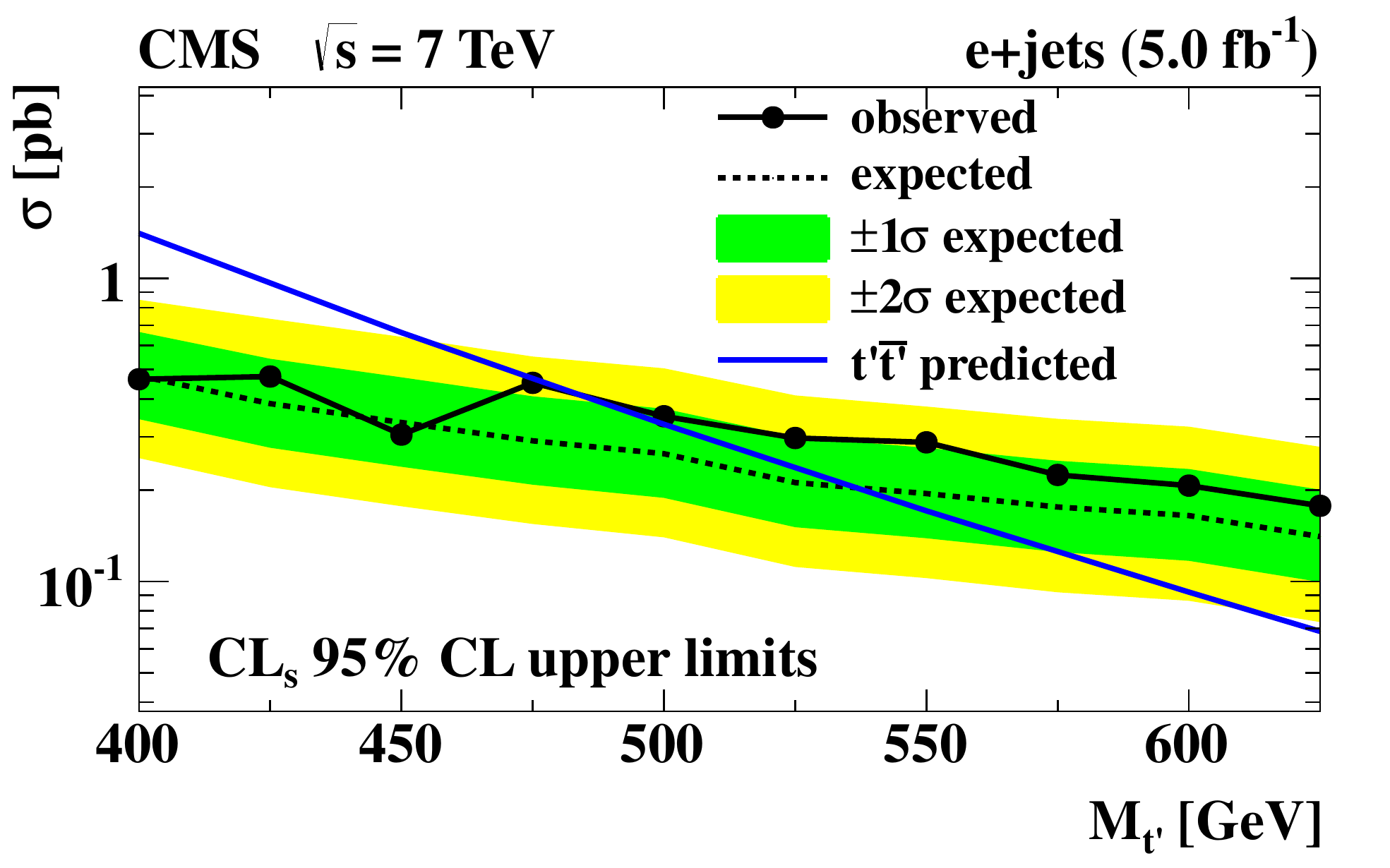}
\includegraphics[width=\cmsFigWidth]{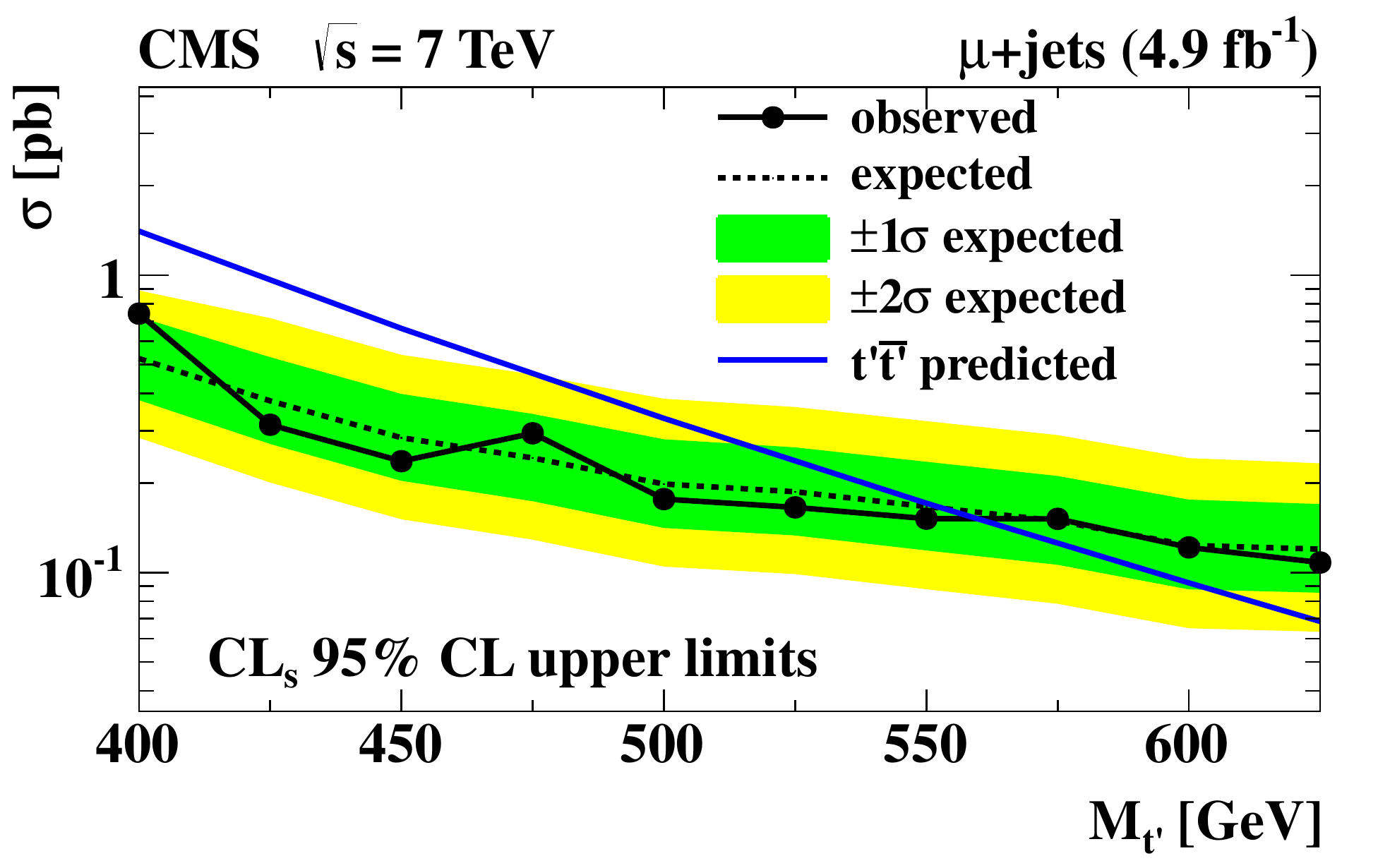}
\includegraphics[width=\cmsFigWidth]{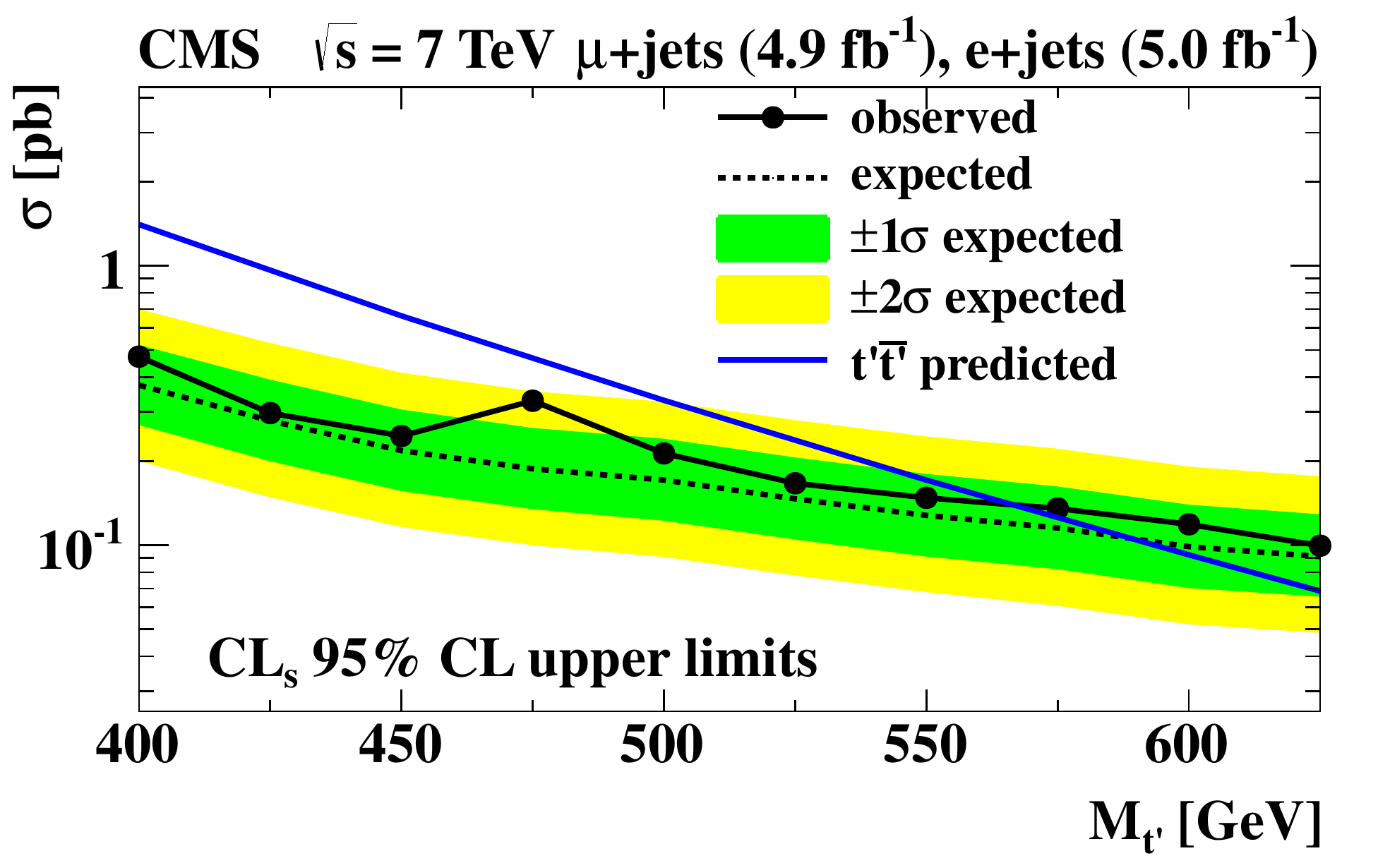}
\caption{The observed (solid line with points) and expected (dotted line) 95\%
\CL{} upper limits on the $\PQtpr\PAQtpr$ production cross section as
a function of the $\PQtpr$-quark mass for $\Pe$+jets (top), $\Pgm$+jets
(middle), and combined (bottom) channels. The ${\pm}1$ and ${\pm}2$ standard deviation
ranges for the expected limits are shown by the bands. The theoretical
$\PQtpr\PAQtpr$ cross section is shown by the continuous line without points.}
\label{fig:limits}
\end{center}
\end{figure}

We estimate the effects of these systematic uncertainties on the expected
$\PQtpr\PAQtpr$ cross section limits by adding them to the limit calculation
one at a time. The largest effects on the expected limits come from the normalizations
of the EW background, the jet energy scale calibration, and the normalization of the
$\ttbar$ background. All other uncertainties change the expected limits
by insignificant amounts. In order to simplify the computational complexity of
the limit computation, we therefore consider only a limited set of systematic
uncertainties in the limit calculation by assigning nuisance parameters to them:
the integrated luminosity, normalization of the EW and $\ttbar$ backgrounds,
lepton efficiency, jet energy scale, and parton-shower matching threshold.
The additional effect of the other uncertainties is negligible.
All of these except the lepton efficiency are treated as correlated in
the combined result from the $\Pe$+jets and the $\Pgm$+jets channels.

\section{Results}
\label{sec:results}

Figure~\ref{fig:limits} shows the observed and expected 95\% \CL{} upper
limits on the $\PQtpr\PAQtpr$ cross section for the $\Pe$+jets
(top), the $\Pgm$+jets (middle) channels, and the combination of both channels
(bottom). The 95\% \CL{} lower limit for the $\PQtpr$-quark mass is given
by the value at which the observed upper limit curve for the $\PQtpr\PAQtpr$
cross section intersects the theoretical curve, also shown in Fig.~\ref{fig:limits}.
In the $\Pe$+jets channel this happens for the 95\% \CL{} observed (expected)
lower limits for a $\PQtpr$-quark mass of 490 (540)\GeV. In the $\Pgm$+jets
channel the corresponding $\PQtpr$-quark mass limit is 560 (550)\GeV. The
combined observed (expected) limit from both channels is 570 (590)\GeV.
A comparable lower limit on the $\PQtpr$ mass of 557\GeV was obtained
recently by the CMS Collaboration using a dilepton channel~\cite{CMS:2012ab}.

\section{Summary}

The results of a search for up-type fourth-generation quarks
that are pair produced in pp interactions at $\sqrt{s}$ = 7\TeV
and decay exclusively to $\PW\cPqb$ have been presented. Events were selected
in which one of the $\PW$ bosons decays to leptons and the other to
a quark-antiquark pair. The selection required an electron or a muon, significant
missing transverse momentum, and at least four jets, of which at least one was
identified as a \cPqb\ jet. A kinematic fit assuming $\PQtpr\PAQtpr$ production
was performed and for every event a candidate $\PQtpr$-quark mass and
the sum over the transverse momenta of all decay products of the
$\PQtpr\PAQtpr$ system were reconstructed. No significant deviations from
the standard model expectations have been found in these two-dimensional
distributions, and upper limits have been set on the production cross section
of such $\PQtpr$ quarks as a function of their mass. By comparing with the
predicted cross section for $\PQtpr\PAQtpr$ production, the strong pair
production of $\PQtpr$ quarks is excluded at 95\% \CL{} for masses below
570\GeV under the model assumptions used in this analysis. This result and
the one from~\cite{CMS:2012ab} are the most restrictive yet found and raise
the lower limit on the mass of a $\PQtpr$ quark to a region where
perturbative calculations for the weak interactions start to fail and
nonperturbative effects become significant. The search is equally sensitive to
nonchiral heavy quarks decaying to $\PW\cPqb$. In this case, the results
can be interpreted as upper limits on the production cross section times the
branching fraction to $\PW\cPqb$.

\section*{Acknowledgements}
We congratulate our colleagues in the CERN accelerator departments for the excellent performance of the LHC machine. We thank the technical and administrative staff at CERN and other CMS institutes, and acknowledge support from: FMSR (Austria); FNRS and FWO (Belgium); CNPq, CAPES, FAPERJ, and FAPESP (Brazil); MES (Bulgaria); CERN; CAS, MoST, and NSFC (China); COLCIENCIAS (Colombia); MSES (Croatia); RPF (Cyprus); MoER, SF0690030s09 and ERDF (Estonia); Academy of Finland, MEC, and HIP (Finland); CEA and CNRS/IN2P3 (France); BMBF, DFG, and HGF (Germany); GSRT (Greece); OTKA and NKTH (Hungary); DAE and DST (India); IPM (Iran); SFI (Ireland); INFN (Italy); NRF and WCU (Korea); LAS (Lithuania); CINVESTAV, CONACYT, SEP, and UASLP-FAI (Mexico); MSI (New Zealand); PAEC (Pakistan); MSHE and NSC (Poland); FCT (Portugal); JINR (Armenia, Belarus, Georgia, Ukraine, Uzbekistan); MON, RosAtom, RAS and RFBR (Russia); MSTD (Serbia); MICINN and CPAN (Spain); Swiss Funding Agencies (Switzerland); NSC (Taipei); TUBITAK and TAEK (Turkey); STFC (United Kingdom); DOE and NSF (USA).

\bibliography{auto_generated}   

\providecommand{\href}[2]{#2}\begingroup\raggedright\begin{thebibliography}{10}%
\makeatletter
\providecommand{\hrefCMSnoop }[0]{\@secondoftwo}%
\makeatother
\providecommand{\doi}{\texttt{doi:}\begingroup \urlstyle{tt}\Url}

\bibitem{Glashow:1961tr}
\hrefCMSnoop {} {S.~L. Glashow, ``{Partial-symmetries of weak interactions}'',}
  \textit{ Nucl. Phys.} \textbf{ 22} (1961) 579,
  \href{http://dx.doi.org/10.1016/0029-5582(61)90469-2}{\doi{10.1016/0029-5582(61)90469-2}}.

\bibitem{PhysRevLett.19.1264}
\hrefCMSnoop {} {S.~Weinberg, ``A Model of Leptons'',} \textit{ Phys. Rev.
  Lett.} \textbf{ 19} (1967) 1264,
  \href{http://dx.doi.org/10.1103/PhysRevLett.19.1264}{\doi{10.1103/PhysRevLett.19.1264}}.

\bibitem{salam}
\hrefCMSnoop {} {A.~Salam, ``Weak and electromagnetic interactions'',} in
  \textit{ Elementary particle physics: relativistic groups and analyticity},
  N.~Svartholm, ed., p.~367.
\newblock Almqvist \& Wiksell, Stockholm, 1968.
\newblock Proceedings of the eighth Nobel symposium.

\bibitem{novikov}
\hrefCMSnoop {} {V.~A. Novikov, A.~N. Rozanov, and M.~I. Vysotsky, ``Once more
  on extra quark-lepton generations and precision measurements'',} \textit{
  Phys. Atom. Nucl.} \textbf{ 73} (2010) 636,
  \href{http://dx.doi.org/10.1134/S1063778810040095}{\doi{10.1134/S1063778810040095}},
  \href{http://www.arXiv.org/abs/0904.4570}{\texttt{ arXiv:0904.4570}}.

\bibitem{Atlas_Higgs}
\hrefCMSnoop {} {{ ATLAS} Collaboration, ``{Observation of a new particle in
  the search for the Standard Model Higgs boson with the ATLAS detector at the
  LHC}'',} \textit{ Phys. Lett. B} \textbf{ 716} (2012) 1,
  \href{http://dx.doi.org/10.1016/j.physletb.2012.08.020}{\doi{10.1016/j.physletb.2012.08.020}},
  \href{http://www.arXiv.org/abs/1207.7214}{\texttt{ arXiv:1207.7214}}.

\bibitem{CMS_Higgs}
\hrefCMSnoop {} {{ CMS} Collaboration, ``{Observation of a new boson at a mass
  of 125 GeV with the CMS experiment at the LHC}'',} \textit{ Phys. Lett. B}
  \textbf{ 716} (2012) 30,
  \href{http://dx.doi.org/10.1016/j.physletb.2012.08.021}{\doi{10.1016/j.physletb.2012.08.021}},
  \href{http://www.arXiv.org/abs/1207.7235}{\texttt{ arXiv:1207.7235}}.

\bibitem{Eberhardt_2012}
\hrefCMSnoop {} {O.~Eberhardt {et~al.}, ``{Joint analysis of Higgs boson decays
  and electroweak precision observables in the standard model with a sequential
  fourth generation}'',} \textit{ Phys. Rev. D} \textbf{ 86} (2012) 013011,
  \href{http://dx.doi.org/10.1103/PhysRevD.86.013011}{\doi{10.1103/PhysRevD.86.013011}},
  \href{http://www.arXiv.org/abs/1204.3872}{\texttt{ arXiv:1204.3872}}.
See also \href{http://www.arxiv.org/abs/1209.1101}{\texttt{arxiv:1209.1101}}.

\bibitem{Chen_He}
\hrefCMSnoop {} {N.~Chen and H.-J. He, ``{LHC signatures of two-Higgs-doublets
  with fourth family}'',} \textit{ JHEP} \textbf{ 04} (2012) 062,
  \href{http://dx.doi.org/10.1007/JHEP04(2012)062}{\doi{10.1007/JHEP04(2012)062}},
  \href{http://www.arXiv.org/abs/1202.3072}{\texttt{ arXiv:1202.3072}}.

\bibitem{beautiful}
\hrefCMSnoop {} {D.~Choudhury, T.~Tait, and C.~Wagner, ``Beautiful Mirrors and
  Precision Electroweak Data'',} \textit{ Phys. Rev. D} \textbf{ 65} (2002)
  053002,
  \href{http://dx.doi.org/10.1103/PhysRevD.65.053002}{\doi{10.1103/PhysRevD.65.053002}},
  \href{http://www.arXiv.org/abs/hep-ph/0109097}{\texttt{
  arXiv:hep-ph/0109097}}.

\bibitem{Schmaltz200340}
\hrefCMSnoop {} {M.~Schmaltz, ``{Physics beyond the standard model (Theory):
  Introducing the Little Higgs}'',} \textit{ Nucl. Phys. Proc. Suppl.} \textbf{
  117} (2003) 40,
  \href{http://dx.doi.org/10.1016/S0920-5632(03)01409-9}{\doi{10.1016/S0920-5632(03)01409-9}},
  \href{http://www.arXiv.org/abs/hep-ph/0210415}{\texttt{
  arXiv:hep-ph/0210415}}.

\bibitem{Kribs}
\hrefCMSnoop {} {G.~D. Kribs {et~al.}, ``{Four generations and Higgs
  physics}'',} \textit{ Phys. Rev. D} \textbf{ 76} (2007) 075016,
  \href{http://dx.doi.org/10.1103/PhysRevD.76.075016}{\doi{10.1103/PhysRevD.76.075016}},
  \href{http://www.arXiv.org/abs/0706.3718}{\texttt{ arXiv:0706.3718}}.

\bibitem{Soni}
\hrefCMSnoop {} {A.~Soni {et~al.}, ``{SM with four generations: Selected
  implications for rare B and K decays}'',} \textit{ Phys. Rev. D} \textbf{ 82}
  (2010) 033009,
  \href{http://dx.doi.org/10.1103/PhysRevD.82.033009}{\doi{10.1103/PhysRevD.82.033009}},
  \href{http://www.arXiv.org/abs/1002.0595}{\texttt{ arXiv:1002.0595}}.

\bibitem{Eberhardt}
\hrefCMSnoop {} {O.~Eberhardt, A.~Lenz, and J.~Rohrwild, ``{Less space for a
  new family of fermions}'',} \textit{ Phys. Rev. D} \textbf{ 82} (2010)
  095006,
  \href{http://dx.doi.org/10.1103/PhysRevD.82.095006}{\doi{10.1103/PhysRevD.82.095006}},
  \href{http://www.arXiv.org/abs/1005.3505}{\texttt{ arXiv:1005.3505}}.

\bibitem{Chanowitz}
\hrefCMSnoop {} {M.~S. Chanowitz, M.~A. Furman, and I.~Hinchliffe, ``Weak
  Interactions of Ultraheavy Fermions'',} \textit{ Phys. Lett. B} \textbf{ 78}
  (1978) 285,
  \href{http://dx.doi.org/10.1016/0370-2693(78)90024-2}{\doi{10.1016/0370-2693(78)90024-2}}.

\bibitem{Chanowitz_CKM}
\hrefCMSnoop {} {M.~S. Chanowitz, ``Bounding {CKM} Mixing with a Fourth
  Family'',} \textit{ Phys. Rev. D} \textbf{ 79} (2009) 113008,
  \href{http://dx.doi.org/10.1103/PhysRevD.79.113008}{\doi{10.1103/PhysRevD.79.113008}},
  \href{http://www.arXiv.org/abs/0904.3570}{\texttt{ arXiv:0904.3570}}.

\bibitem{D0}
\hrefCMSnoop {} {{ D0} Collaboration, ``Search for a Fourth Generation $t'$
  Quark in $p\bar{p}$ Collisions at $\sqrt{s}=1.96$~{TeV}'',} \textit{ Phys.
  Rev. Lett.} \textbf{ 107} (2011) 082001,
  \href{http://dx.doi.org/10.1103/PhysRevLett.107.082001}{\doi{10.1103/PhysRevLett.107.082001}},
  \href{http://www.arXiv.org/abs/1104.4522}{\texttt{ arXiv:1104.4522}}.

\bibitem{CDF}
\hrefCMSnoop {} {{ CDF} Collaboration, ``Search for a Heavy Toplike Quark in
  $p\bar{p}$ Collisions at $\sqrt{s}=1.96$~{TeV}'',} \textit{ Phys. Rev. Lett.}
  \textbf{ 107} (2011) 261801,
  \href{http://dx.doi.org/10.1103/PhysRevLett.107.261801}{\doi{10.1103/PhysRevLett.107.261801}},
  \href{http://www.arXiv.org/abs/1107.3875}{\texttt{ arXiv:1107.3875}}.

\bibitem{ATLAS_1lepton}
\hrefCMSnoop {} {{ ATLAS} Collaboration, ``Search for Pair Production of a
  Heavy Up-Type Quark Decaying to a {W} Boson and a b Quark in the lepton+jets
  Channel with the {ATLAS} Detector'',} \textit{ Phys. Rev. Lett.} \textbf{
  108} (2012) 261802,
  \href{http://dx.doi.org/10.1103/PhysRevLett.108.261802}{\doi{10.1103/PhysRevLett.108.261802}},
\href{http://www.arXiv.org/abs/1202.3076}{\texttt{ arXiv:1202.3076}}.

\bibitem{CMS}
\hrefCMSnoop {} {{ CMS} Collaboration, ``{The CMS experiment at the CERN
  LHC}'',} \textit{ JINST} \textbf{ 3} (2008) S08004,
\href{http://dx.doi.org/10.1088/1748-0221/3/08/S08004}{\doi{10.1088/1748-0221/3/08/S08004}}.

\bibitem{powheg1}
\hrefCMSnoop {} {{P. Nason}, ``{A new method for combining NLO QCD with shower
  Monte Carlo algorithms}'',} \textit{ JHEP} \textbf{ 11} (2004) 040,
  \href{http://dx.doi.org/10.1088/1126-6708/2004/11/040}{\doi{10.1088/1126-6708/2004/11/040}},
  \href{http://www.arXiv.org/abs/hep-ph/0409146}{\texttt{
  arXiv:hep-ph/0409146}}.

\bibitem{powheg2}
\hrefCMSnoop {} {S.~Frixione, P.~Nason, and C.~Oleari, ``{Matching NLO QCD
  computations with parton shower simulations: the POWHEG method}'',} \textit{
  JHEP} \textbf{ 11} (2007) 070,
  \href{http://dx.doi.org/10.1088/1126-6708/2007/11/070}{\doi{10.1088/1126-6708/2007/11/070}},
  \href{http://www.arXiv.org/abs/0709.2092}{\texttt{ arXiv:0709.2092}}.

\bibitem{powheg3}
\hrefCMSnoop {} {S.~Alioli {et~al.}, ``{A general framework for implementing
  NLO calculations in shower Monte Carlo programs: the POWHEG BOX}'',} \textit{
  JHEP} \textbf{ 06} (2010) 043,
  \href{http://dx.doi.org/10.1007/JHEP06(2010)043}{\doi{10.1007/JHEP06(2010)043}},
  \href{http://www.arXiv.org/abs/1002.2581}{\texttt{ arXiv:1002.2581}}.

\bibitem{madgraph}
\hrefCMSnoop {} {J.~Alwall {et~al.}, ``{MadGraph/MadEvent v4: the new web
  generation}'',} \textit{ JHEP} \textbf{ 09} (2007) 028,
  \href{http://dx.doi.org/10.1088/1126-6708/2007/09/028}{\doi{10.1088/1126-6708/2007/09/028}},
  \href{http://www.arXiv.org/abs/0706.2334}{\texttt{ arXiv:0706.2334}}.

\bibitem{pythia}
\hrefCMSnoop {} {T.~Sj$\stackrel{..}{\text o}$strand, S.~Mrenna, and P.~Z.
  Skands, ``{PYTHIA 6.4 physics and manual}'',} \textit{ JHEP} \textbf{ 05}
  (2006) 026,
  \href{http://dx.doi.org/10.1088/1126-6708/2006/05/026}{\doi{10.1088/1126-6708/2006/05/026}},
  \href{http://www.arXiv.org/abs/hep-ph/0603175}{\texttt{
  arXiv:hep-ph/0603175}}.

\bibitem{geant4}
\hrefCMSnoop {} {{ GEANT4} Collaboration, ``{GEANT4 -- a simulation
  toolkit}'',} \textit{ Nucl. Instrum. Meth. A} \textbf{ 506} (2003) 250,
  \href{http://dx.doi.org/10.1016/S0168-9002(03)01368-8}{\doi{10.1016/S0168-9002(03)01368-8}}.

\bibitem{particleflow}
\href {https://cdsweb.cern.ch/record/1194487} {{CMS Collaboration},
  ``Particle-Flow Event Reconstruction in {CMS} and Performance for Jets, Taus,
  and {MET}'',} CMS Physics Analysis Summary CMS-PAS-PFT-09-001, (2009).

\bibitem{particleflow1}
\href {https://cdsweb.cern.ch/record/1279341} {{CMS Collaboration},
  ``Commissioning of the Particle-Flow Reconstruction in Minimum-Bias and Jet
  Events from pp Collisions at 7 {TeV}'',} CMS Physics Analysis Summary
  CMS-PAS-PFT-10-002, (2010).

\bibitem{particleflow2}
\href {https://cdsweb.cern.ch/record/1279347} {{CMS Collaboration},
  ``Particle-flow commissioning with muons and electrons from {J/Psi} and {W}
  events at 7 {TeV}'',} CMS Physics Analysis Summary CMS-PAS-PFT-10-003,
  (2010).

\bibitem{antikt}
\hrefCMSnoop {} {M.~Cacciari, G.~P. Salam, and G.~Soyez, ``{The anti-$k_t$ jet
  clustering algorithm}'',} \textit{ JHEP} \textbf{ 04} (2008) 063,
  \href{http://dx.doi.org/10.1088/1126-6708/2008/04/063}{\doi{10.1088/1126-6708/2008/04/063}},
  \href{http://www.arXiv.org/abs/0802.1189}{\texttt{ arXiv:0802.1189}}.

\bibitem{Cacciari:2005hq}
\hrefCMSnoop {} {M.~Cacciari and G.~P. Salam, ``{Dispelling the $N^{3}$ myth
  for the $k_t$ jet-finder}'',} \textit{ Phys. Lett. B} \textbf{ 641} (2006)
  57,
  \href{http://dx.doi.org/10.1016/j.physletb.2006.08.037}{\doi{10.1016/j.physletb.2006.08.037}},
  \href{http://www.arXiv.org/abs/hep-ph/0512210}{\texttt{
  arXiv:hep-ph/0512210}}.

\bibitem{fastjet1}
\hrefCMSnoop {} {M.~Cacciari, G.~P. Salam, and G.~Soyez, ``{The catchment area
  of jets}'',} \textit{ JHEP} \textbf{ 04} (2008) 005,
  \href{http://dx.doi.org/10.1088/1126-6708/2008/04/005}{\doi{10.1088/1126-6708/2008/04/005}},
  \href{http://www.arXiv.org/abs/0802.1188}{\texttt{ arXiv:0802.1188}}.

\bibitem{fastjet2}
\hrefCMSnoop {} {M.~Cacciari and G.~P. Salam, ``{Pileup subtraction using jet
  areas}'',} \textit{ Phys. Lett. B} \textbf{ 659} (2008) 119,
  \href{http://dx.doi.org/10.1016/j.physletb.2007.09.077}{\doi{10.1016/j.physletb.2007.09.077}},
  \href{http://www.arXiv.org/abs/0707.1378}{\texttt{ arXiv:0707.1378}}.

\bibitem{Chatrchyan:2011ds}
\hrefCMSnoop {} {{ CMS} Collaboration, ``{Determination of jet energy
  calibration and transverse momentum resolution in CMS}'',} \textit{ JINST}
  \textbf{ 6} (2011) P11002,
  \href{http://dx.doi.org/10.1088/1748-0221/6/11/P11002}{\doi{10.1088/1748-0221/6/11/P11002}},
  \href{http://www.arXiv.org/abs/1107.4277}{\texttt{ arXiv:1107.4277}}.

\bibitem{CMS-PAS-BTV-11-003}
\href {http://cdsweb.cern.ch/record/1421611} {{CMS Collaboration},
  ``Measurement of the $b$-tagging efficiency using $t\bar{t}$ events'',} CMS
  Physics Analysis Summary CMS-PAS-BTV-11-003, (2011).

\bibitem{Chatrchyan:2011yy}
\hrefCMSnoop {} {{ CMS} Collaboration, ``{Measurement of the
  $\mathrm{t\bar{t}}$ production cross section in pp collisions at 7\TeV in
  lepton + jets events using b-quark jet identification}'',} \textit{ Phys.
  Rev. D} \textbf{ 84} (2011) 092004,
  \href{http://dx.doi.org/10.1103/PhysRevD.84.092004}{\doi{10.1103/PhysRevD.84.092004}},
  \href{http://www.arXiv.org/abs/1108.3773}{\texttt{ arXiv:1108.3773}}.

\bibitem{MCFM}
\hrefCMSnoop {} {J.~M. Campbell and R.~Ellis, ``{Next-to-leading order
  corrections to $W + $ 2 jet and $Z + $ 2 jet production at hadron
  colliders}'',} \textit{ Phys. Rev. D} \textbf{ 65} (2002) 113007,
  \href{http://dx.doi.org/10.1103/PhysRevD.65.113007}{\doi{10.1103/PhysRevD.65.113007}},
  \href{http://www.arXiv.org/abs/hep-ph/0202176}{\texttt{
  arXiv:hep-ph/0202176}}.

\bibitem{HATHOR}
\hrefCMSnoop {} {M.~Aliev {et~al.}, ``{HATHOR -- HAdronic Top and Heavy quarks
  crOss section calculatoR}'',} \textit{ Comput. Phys. Commun.} \textbf{ 182}
  (2011) 1034,
  \href{http://dx.doi.org/10.1016/j.cpc.2010.12.040}{\doi{10.1016/j.cpc.2010.12.040}},
\href{http://www.arXiv.org/abs/1007.1327}{\texttt{ arXiv:1007.1327}}.

\bibitem{CLs1}
\hrefCMSnoop {} {T.~Junk, ``{Confidence level computation for combining
  searches with small statistics}'',} \textit{ Nucl. Instrum. Meth. A} \textbf{
  434} (1999) 435,
  \href{http://dx.doi.org/10.1016/S0168-9002(99)00498-2}{\doi{10.1016/S0168-9002(99)00498-2}},
  \href{http://www.arXiv.org/abs/hep-ex/9902006}{\texttt{
  arXiv:hep-ex/9902006}}.

\bibitem{CLs2}
\hrefCMSnoop {} {A.~L. Read, ``{Presentation of search results: The $CL_s$
  technique}'',} \textit{ J. Phys. G} \textbf{ 28} (2002) 2693,
  \href{http://dx.doi.org/10.1088/0954-3899/28/10/313}{\doi{10.1088/0954-3899/28/10/313}}.

\bibitem{asymptoticCLs}
\hrefCMSnoop {} {G.~Cowan {et~al.}, ``{Asymptotic formulae for likelihood-based
  tests of new physics}'',} \textit{ Eur. Phys. J. C} \textbf{ 71} (2011) 1554,
  \href{http://dx.doi.org/10.1140/epjc/s10052-011-1554-0}{\doi{10.1140/epjc/s10052-011-1554-0}},
  \href{http://www.arXiv.org/abs/1007.1727}{\texttt{ arXiv:1007.1727}}.

\bibitem{luminosity}
\href {http://cdsweb.cern.ch/record/1434360} {{CMS Collaboration}, ``Absolute
  Calibration of the Luminosity Measurement at {CMS}: Winter 2012 Update'',}
  CMS Physics Analysis Summary CMS-PAS-SMP-2012-008, (2012).

\bibitem{matching}
\hrefCMSnoop {} {S.~Mrenna and P.~Richardson, ``{Matching matrix elements and
  parton showers with HERWIG and PYTHIA}'',} \textit{ JHEP} \textbf{ 05} (2004)
  040,
  \href{http://dx.doi.org/10.1088/1126-6708/2004/05/040}{\doi{10.1088/1126-6708/2004/05/040}},
  \href{http://www.arXiv.org/abs/hep-ph/0312274}{\texttt{
  arXiv:hep-ph/0312274}}.

\bibitem{CMS:2012ab}
\hrefCMSnoop {} {{ CMS} Collaboration, ``{Search for heavy, top-like quark pair
  production in the dilepton final state in pp collisions at $\sqrt{s}$ = 7
  TeV}'',} \textit{ Phys. Lett. B} \textbf{ 716} (2012) 103,
  \href{http://dx.doi.org/10.1016/j.physletb.2012.07.059}{\doi{10.1016/j.physletb.2012.07.059}},
  \href{http://www.arXiv.org/abs/1203.5410}{\texttt{ arXiv:1203.5410}}.

\end{thebibliography}\endgroup
\cleardoublepage \appendix\section{The CMS Collaboration \label{app:collab}}\begin{sloppypar}\hyphenpenalty=5000\widowpenalty=500\clubpenalty=5000\textbf{Yerevan Physics Institute,  Yerevan,  Armenia}\\*[0pt]
S.~Chatrchyan, V.~Khachatryan, A.M.~Sirunyan, A.~Tumasyan
\vskip\cmsinstskip
\textbf{Institut f\"{u}r Hochenergiephysik der OeAW,  Wien,  Austria}\\*[0pt]
W.~Adam, E.~Aguilo, T.~Bergauer, M.~Dragicevic, J.~Er\"{o}, C.~Fabjan\cmsAuthorMark{1}, M.~Friedl, R.~Fr\"{u}hwirth\cmsAuthorMark{1}, V.M.~Ghete, J.~Hammer, N.~H\"{o}rmann, J.~Hrubec, M.~Jeitler\cmsAuthorMark{1}, W.~Kiesenhofer, V.~Kn\"{u}nz, M.~Krammer\cmsAuthorMark{1}, I.~Kr\"{a}tschmer, D.~Liko, I.~Mikulec, M.~Pernicka$^{\textrm{\dag}}$, B.~Rahbaran, C.~Rohringer, H.~Rohringer, R.~Sch\"{o}fbeck, J.~Strauss, A.~Taurok, W.~Waltenberger, G.~Walzel, C.-E.~Wulz\cmsAuthorMark{1}
% ...
\vskip\cmsinstskip
\dag:~Deceased\\
1:~~Also at Vienna University of Technology, Vienna, Austria\\
% ...

\end{sloppypar}
\end{document}